\documentclass[referee]{raa}            
\usepackage{graphicx,times}           
\usepackage{natbib}
\usepackage{amssymb,amsmath}
\bibpunct{(}{)}{;}{a}{}{,}
\usepackage[pagebackref=true]{hyperref}
\usepackage{caption}
\usepackage{booktabs}
\begin{document}  
  \title{Noise Reduction Method for Radio Astronomy Single Station Observation Based on Wavelet Transform and Mathematical Morphology}
   \volnopage{Vol.0 (20xx) No.0, 000--000}      
   \setcounter{page}{1}         
   \author{Ming-wei Qin
      \inst{1}\textasteriskcentered
   \and Rui Tang
      \inst{1}
   \and Ying-hui Zhou
      \inst{1}
    \and Chang-jun Lan
      \inst{1}
    \and Wen-hao Fu
      \inst{1}
    \and Huan Wang
      \inst{1}
    \and Bao-lin Hou
      \inst{1}
    \and Zamri
      \inst{2}
    \and Jin-song Ping
      \inst{3}\textasteriskcentered
    \and Wen-jun Yang
      \inst{4}
    \and Liang Dong
      \inst{5}
   }

   \institute{ School of Information Engineering, Southwest University of Science and Technology, Mianyang 621010, China;{\it qinmingwei@swust.edu.cn}\\
    \and
             Department of Physics, Lingkungan Budi, University Malaya, 50603, Kuala Lumpur;\\
    \and National Astronomical Observatories, Chinese Academy of Sciences,
             Beijing 100012, China; {\it jsping@bao.ac.cn}\\
    \and
             Xinjiang Astronomical Observatory, Chinese Academy of Sciences, Xinjiang 830011, China;\\
    \and
            Yunnan Astronomical Observatory, Chinese Academy of Sciences, Yunnan 650216, China;\\
\vs\no
   {\small Received 20xx month day; accepted 20xx month day}}

\abstract{The 21 cm radiation of neutral hydrogen provides crucial information for studying the early universe and its evolution. To advance this research, countries have made significant investments in constructing large low-frequency radio telescope arrays, such as the Low Frequency Array (LOFAR) and the Square Kilometre Array Phase 1 Low Frequency (SKA1-low). These instruments are pivotal for radio astronomy research. However, challenges such as ionospheric plasma interference, ambient radio noise, and instrument-related effects have become increasingly prominent, posing major obstacles in cosmology research. To address these issues, this paper proposes an efficient signal processing method that combines wavelet transform and mathematical morphology. The method involves the following steps: Background Subtraction: Background interference in radio observation signals is eliminated. Wavelet Transform: The signal, after removing background noise, undergoes a two-dimensional discrete wavelet transform. Threshold processing is then applied to the wavelet coefficients to effectively remove interference components. Wavelet Inversion: The processed signal is reconstructed using wavelet inversion. Mathematical Morphology: The reconstructed signal is further optimized using mathematical morphology to refine the results. Experimental verification was conducted using solar observation data from the Xinjiang Observatory and the Yunnan Observatory. The results demonstrate that this method successfully removes interference signals while preserving useful signals, thus improving the accuracy of radio astronomy observations and reducing the impact of radio frequency interference (RFI).
\keywords{Radio astronomy; Radio frequency interference; Wavelet transform; Mathematical morphology; Image processing}
}

   \authorrunning{Mingwei Qin, Rui Tang et al }            
  \titlerunning{Photometry of $\delta$ Sct and Related Stars (I) }  
   \maketitle

\section{Introduction}       
\subsection{Background}
The emergence of 21-cm cosmology marked a significant turning point in the fields of astronomy and cosmology \citealt{1}). The 21 cm radiation of neutral hydrogen, which arises from the spin-flip transition of electrons and protons in hydrogen atoms, provides crucial insights into the early universe. In particular, it sheds light on the evolution of the universe during the "Dark Ages" and the "Age of Reionization" (\citealt{2,3,4}). By observing 21 cm radiation at different redshifts, scientists can map the distribution of neutral hydrogen across various epochs of the universe. This enables them to study the formation of the first generation of stars and galaxies, trace the process of gas reionization, and precisely measure the early evolution of the large-scale structure of the universe. These observations reveal the transition of the universe from darkness to light, offering a vivid narrative of its early history (\citealt{5,6,7}). To facilitate these studies, numerous large low-frequency radio telescope arrays have been constructed worldwide. Notable examples include 21CMA (\citealt{8}), LOFAR (\citealt{9}), MWA (\citealt{10}), PAPER (\citealt{11}), HERA (\citealt{12}) and the next-generation large low-frequency array SKA1-low (\citealt{13}), which is under construction. Using large-scale array technology, these telescopes can cover a wider frequency range and significantly improve the signal-to-noise ratio, resulting in higher-resolution images. Using advanced large-scale array technology, these telescopes can cover a broader frequency range, significantly enhance the signal-to-noise ratio, and produce higher-resolution images. These concerted efforts have not only advanced the field of 21 cm cosmology but have also provided a powerful tool for uncovering the universe's underlying structure and evolution.\newline
However, as an open broadband system, astronomical observations face significant challenges from RFI. Many observation frequency bands are not specifically allocated for radio astronomy services by the International Telecommunication Union, making them particularly susceptible to interference from human-generated signals (\citealt{14}). This issue is compounded by various factors along the electromagnetic wave propagation path. These include plasma interference from the Earth's ionosphere, background radio noise from the surrounding environment of the observation station, the effects of the observation instruments themselves, and other unrelated electromagnetic signals. These sources of RFI severely impact the accurate detection of cosmic radio signals. The presence of RFI not only limits the telescope's ability to produce high-resolution imaging but also imposes significant constraints on the scientific applications of the signals. These applications include studies of stellar activity and space weather prediction (\citealt{15,16,17}). Effectively suppressing such interference in complex electromagnetic environments has therefore become a key focus in current research.\newline
With the rapid development of global radio communication, researchers worldwide have proposed various methods and models to address the increasingly serious problem of RFI. Among these, the wavelet transform, known for its excellent time-frequency localization capabilities, has been widely applied in the suppression of RFI. This technique effectively distinguishes signals from noise (\citealt{18,19}). However, the number of decomposition layers used in the wavelet transform plays a crucial role in the final interference suppression outcome. Using too many or too few layers can result in signal distortion or inadequate interference elimination. In contrast, mathematical morphology analyzes the structure of signals using morphological operations, making it effective for removing interference components of specific frequencies (\citealt{20}). However, this method may lead to a loss of signal detail during processing. As a result, there is a strong practical need to develop fast and efficient noise reduction methods that can suppress interference while preserving the integrity of useful signals.
\subsection{Related Work}
In recent years, various measures have been developed to mitigate and suppress RFI in single-station observations. These methods can be broadly categorized into traditional techniques and artificial intelligence-based approaches. Among these methods, (\citealt{21}) applied singular value decomposition (SVD) to the time-frequency matrix. By identifying and reconstructing the matrix using the largest singular value of the offset zero, they effectively mitigated the impact of the RFI. 
(\citealt{22}) employed independent component analysis (ICA) to decompose mixed signals into independent components. Using statistical characteristics, they were able to effectively identify and suppress interference. 
(\citealt{23}) introduced linear prediction technology to recover lost spectral components in the original data. They also implemented a two-step notch filter with limited bandwidth to effectively reduce the impact of interference.  
(\citealt{24}) achieved effective noise reduction by developing an NBI model and estimating its parameters based on the characteristics of RFI.
(\citealt{25}) proposed a fast, sparse method to eliminate RFI outside the main frequency band. (\citealt{26}) introduced an operator modeling method based on an echo dictionary and proposed a dictionary-based suppression algorithm within the framework of robust principal component analysis.
In addition, (\citealt{27}) combined interferometric synthetic aperture radar (SAR) data pairs with RFI models to reconstruct SAR images using a combined low-rank and sparse optimization method.\newline
The use of deep learning-based methods for the suppression of RFI has also shown promising results. 
(\citealt{28}) proposed a new method using convolutional neural network (CNN) technology to reduce the impact of RFI in radio data. The study employed the U-Net framework, which can identify and locate clean signals as well as RFI in two-dimensional time series data obtained from radio telescopes. 
(\citealt{29}) proposed a robust CNN model for identifying RFI using machine learning methods.
(\citealt{30}) proposed a deep fully convolutional neural network (DFCN) that utilizes both amplitude and phase information from interference data to identify RFI.\newline
However, deep learning-based approaches face challenges such as the need for large amounts of labeled data, high computational resource consumption, and difficulties in real-time processing. Most traditional processing methods are based on well-established signal processing theories, which have been validated through long-term practice. These methods are highly stable and can reliably handle a variety of RFI scenarios. However, when it comes to processing large amounts of data over extended periods, some traditional algorithms are relatively slow. Therefore, this paper applies the wavelet transform and digital morphology to radio signal noise reduction, aiming to develop a fast and efficient method for RFI suppression.
\subsection{Solution and Contributions of This Article}
To address the problem of RFI, this study takes advantage of the characteristics of RFI in the radio spectrum and proposes an innovative suppression method that combines the strengths of wavelet transform and mathematical morphology. First, background noise in the radio spectrum is effectively removed through data pre-processing, laying a solid foundation for subsequent analysis. Next, three layers of two-dimensional discrete wavelet decomposition are applied for multi-scale analysis of the preprocessed data. The high-frequency wavelet coefficients obtained after decomposition are thresholded, and the absolute median threshold is used to detect outliers in the low-frequency coefficients. Finally, the wavelet-reconstructed signal is processed using mathematical morphology to eliminate isolated and discontinuous RFI, further enhancing the signal quality. This method achieves fast and efficient RFI suppression while effectively preserving useful signals. Experimental results based on measured data validate the performance and potential of this method, demonstrating its wide application prospects in complex electromagnetic environments.\newline
The main contributions of this paper are as follows.\newline
1. To address the limitations of traditional RFI mitigation methods in protecting useful signals, this paper proposes an innovative processing approach that combines the wavelet transform and mathematical morphology for effective RFI suppression. The process emphasizes the separation and protection of useful signals, ensuring that the integrity and quality of the signals are preserved as much as possible while suppressing interference. Additionally, the paper analyzes the similar characteristics of RFI between different stations, demonstrating that the proposed method provides excellent noise reduction results in various scenarios. This finding not only improves the applicability of the method but also offers a new perspective for developing a general RFI suppression strategy.\newline
2. Based on the characteristics of RFI in low-frequency observations, this paper proposes an RF signal suppression method that preserves useful signals. The method consists of three key steps: data preprocessing, two-dimensional discrete wavelet transform, and mathematical morphology processing. Compared to traditional suppression techniques, the proposed method not only effectively reduces RFI but also prioritizes the preservation of image details. This approach helps to avoid the difficulty of image interpretation caused by the excessive loss of useful signals.
\subsection{Organization of This Article}
The rest of this paper is organized as follows: Section 2 introduces the mathematical model of RFI signals. Section 3 provides a detailed description of the noise reduction workflow using the two-dimensional discrete wavelet transform and mathematical morphology. Section 4 presents the experimental results and performance analysis of the proposed method. Finally, Section 5 concludes the paper and discusses the potential future applications of this method.
\section{Signal Model of RFI}
\begin{figure}[h]
  \begin{minipage}[t]{0.495\linewidth}
  \centering
   \includegraphics[width=60mm]{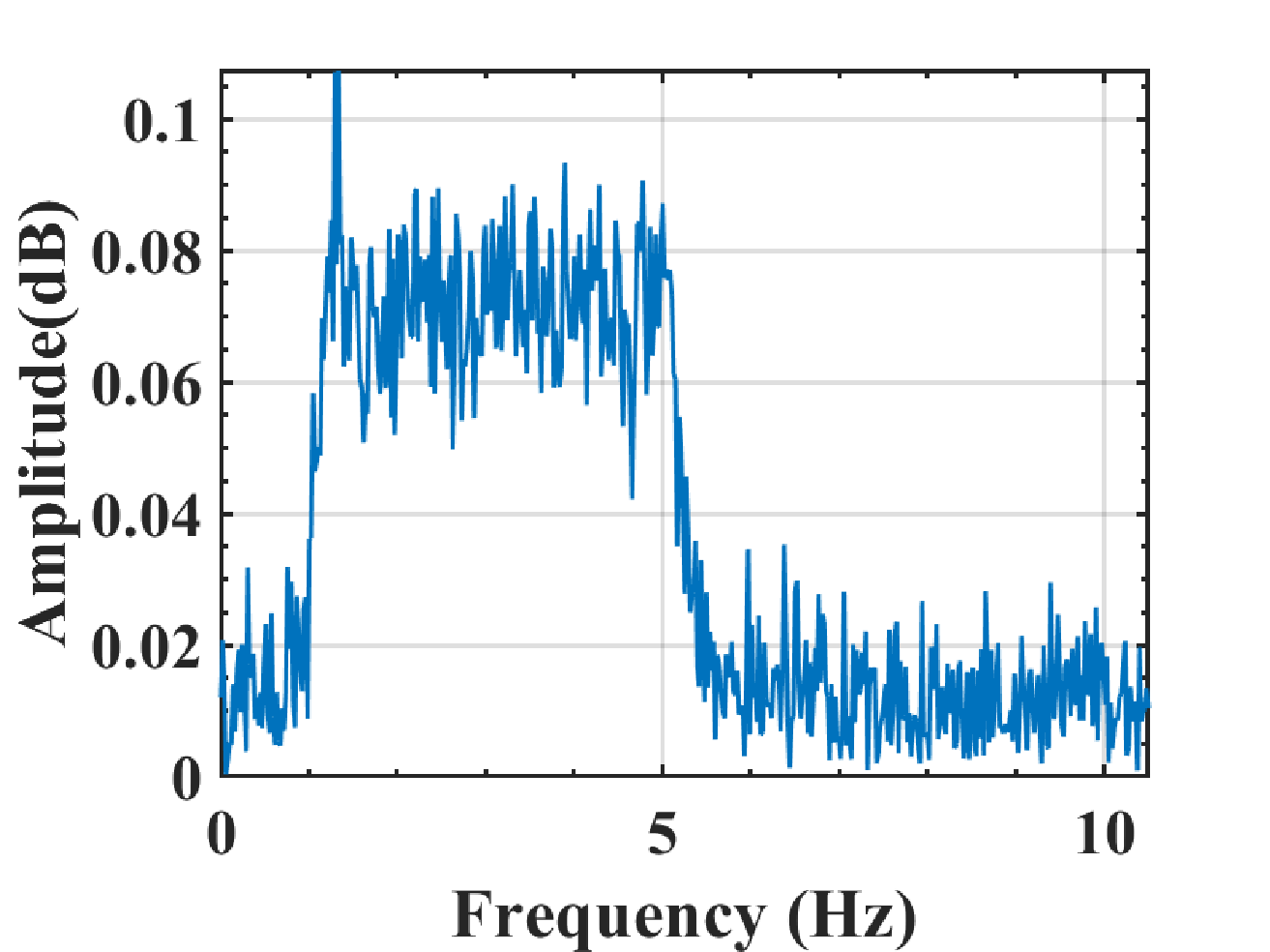}
	  \caption{\label{Fig1}{\small WBI in the frequency domain.} }
  \end{minipage}
  \begin{minipage}[t]{0.495\textwidth}
  \centering
   \includegraphics[width=60mm]{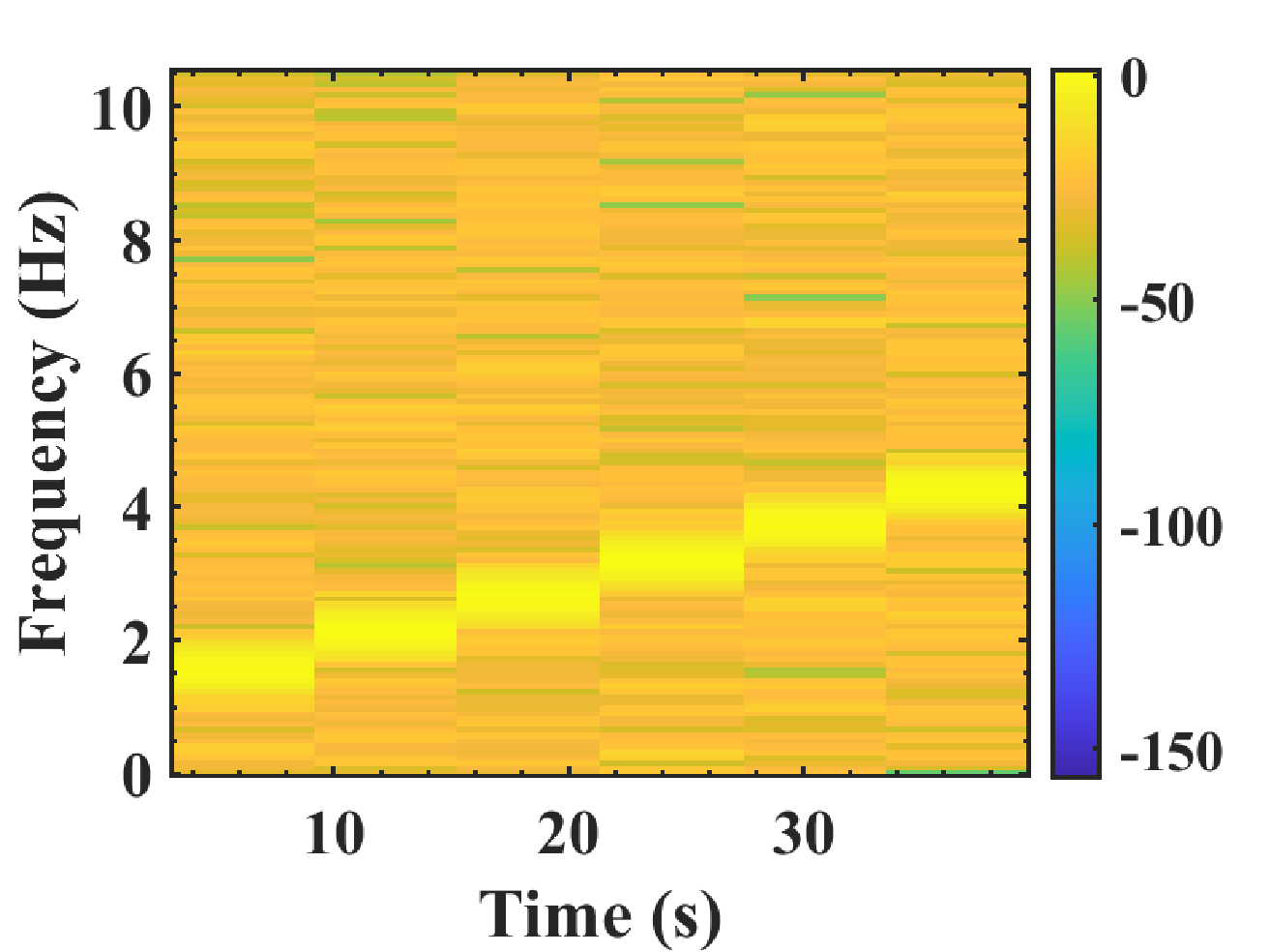}
	  \caption{\label{Fig2}{\small WBI time-frequency spectrogram.}}
  \end{minipage}
    \begin{minipage}[t]{0.495\linewidth}
  \centering
   \includegraphics[width=60mm]{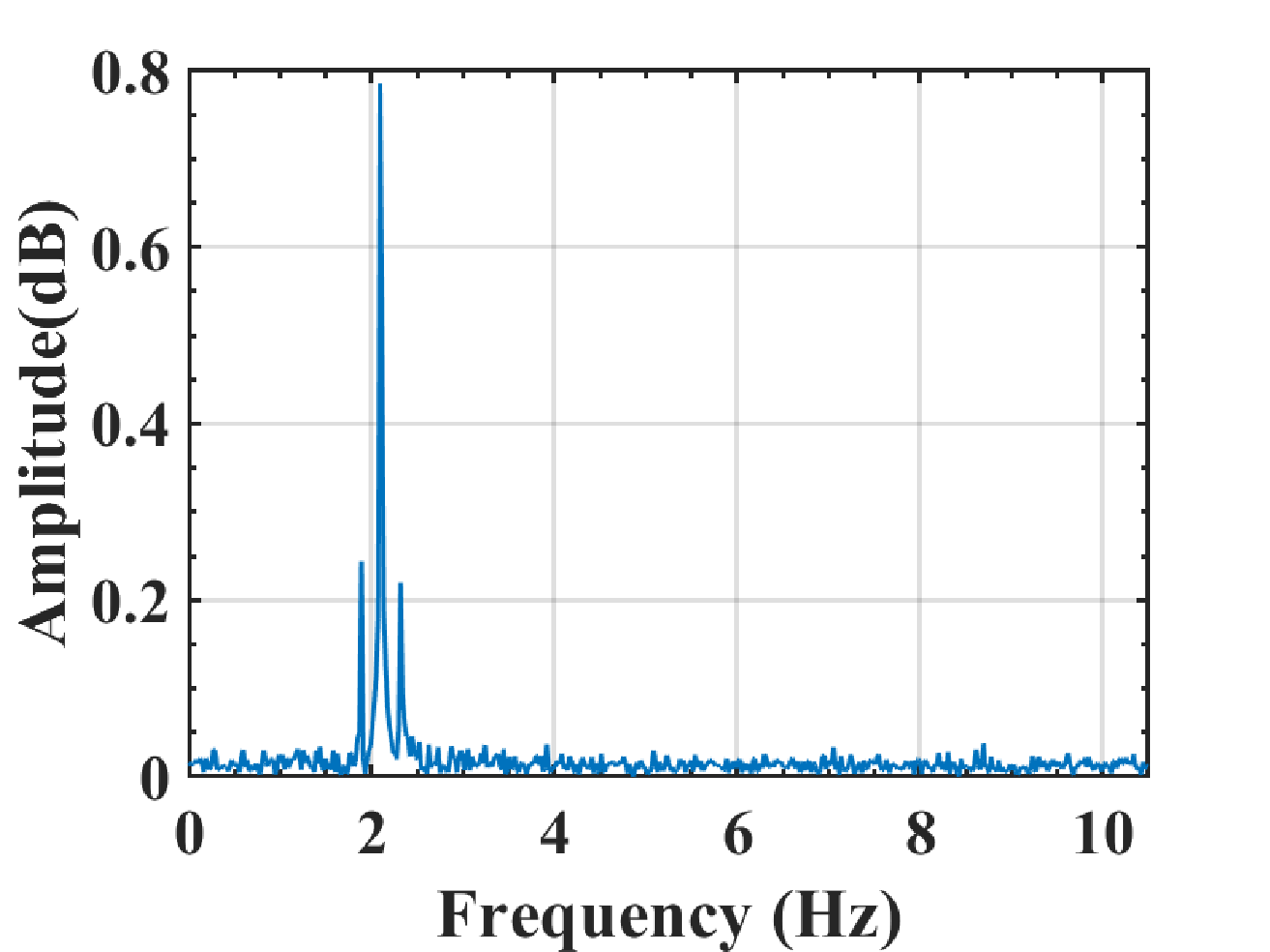}
	  \caption{\label{Fig3}{\small NBI in the frequency domain.} }
  \end{minipage}
  \begin{minipage}[t]{0.495\textwidth}
  \centering
   \includegraphics[width=60mm]{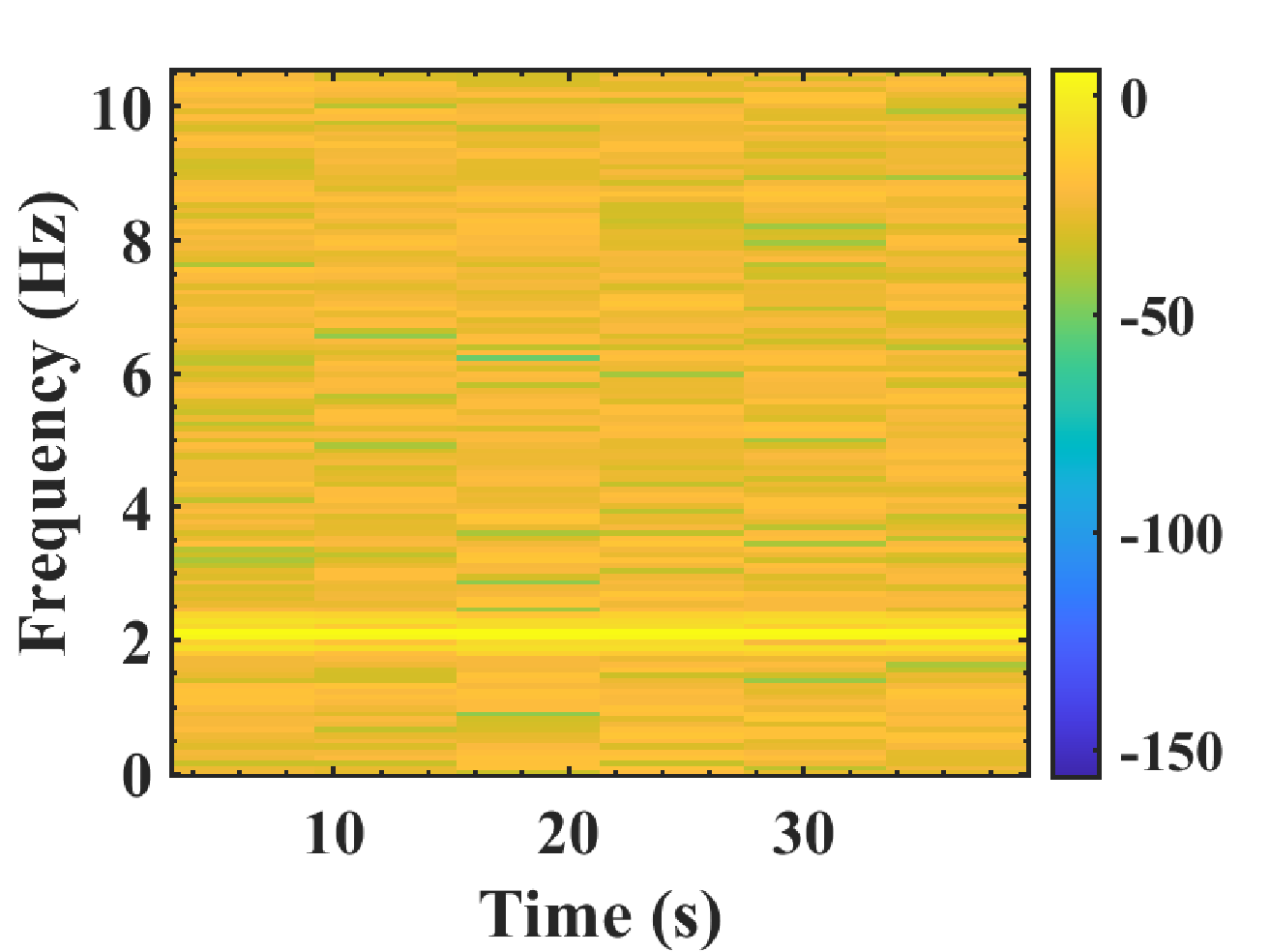}
	  \caption{\label{Fig4}{\small NBI time-frequency spectrogram.}}
  \end{minipage}
\end{figure}

Since radio astronomy observations are often affected by significant RFI, the primary source of interference is the non-correlated electromagnetic signals generated by human activities within the same frequency band as the radio observations. When a radio telescope is in operation, the radio signal it receives can be modeled as shown in \autoref{eq1}:
\begin{equation}\label{eq1}
  \mathrm{y(n)=x(n)+\omega(n)+\varphi(n)}
\end{equation}
Where $x(n)$ represents the actual radio signal, $\omega(n)$ denotes the inherent noise of the system, and $\varphi(n)$ is the RFI signal. For the received signal $y(n)$, $n$ is the sampling index.\newline
Although RFI can take many different forms, radio astronomy observations are generally classified as wideband interference (WBI) and narrowband interference (NBI) (\citealt{31}). 
WBI is typically caused by sudden events such as lightning strikes, short circuits in electronic devices, or issues with Ethernet cables. Its spectral characteristics include broad frequency coverage and high signal strength, as shown in (Fig.~\ref{Fig1}). In the time-frequency spectrogram, it appears as a bright stripe, as shown in (Fig.~\ref{Fig2}).
The mathematical expression for this interference is given in the following \autoref{eq2}:
\begin{equation}\label{eq2}
  I_{\mathrm{BB}}=\sum_{l=1}^{L} A_{l} \exp \left[j\left(2 \pi f_{l} n+\theta_{l}\right)\right]
\end{equation}
Where $L$ is the number of RFI sources, and $A_l$, $f_l$, and $\theta_l$ represent the amplitude, frequency, and phase of the first interference component of WBI, respectively.\newline
During data acquisition, radio telescopes may encounter NBI from sources such as radio stations, mobile communications, and other transmitters. These disturbances appear as distinct frequency spikes in the frequency domain, typically concentrated within a specific frequency range, as shown in (Fig.~\ref{Fig3}) of the spectrum diagram. In the time-frequency spectrogram, NBI is represented as a horizontal line, as shown in (Fig.~\ref{Fig4}). The mathematical expression for this interference is given in \autoref{eq3}:
\begin{equation}\label{eq3}
  I_{\mathrm{NB}}=A \exp [j(2 \pi f n+\theta)]
\end{equation}
Where $j$ is the imaginary unit, and $A$, $f$, and $\theta$ represent the amplitude, frequency, and initial phase of NBI, respectively.
\section{Methodology}
\begin{figure}
   \centering
   \includegraphics[width=\textwidth, angle=0]{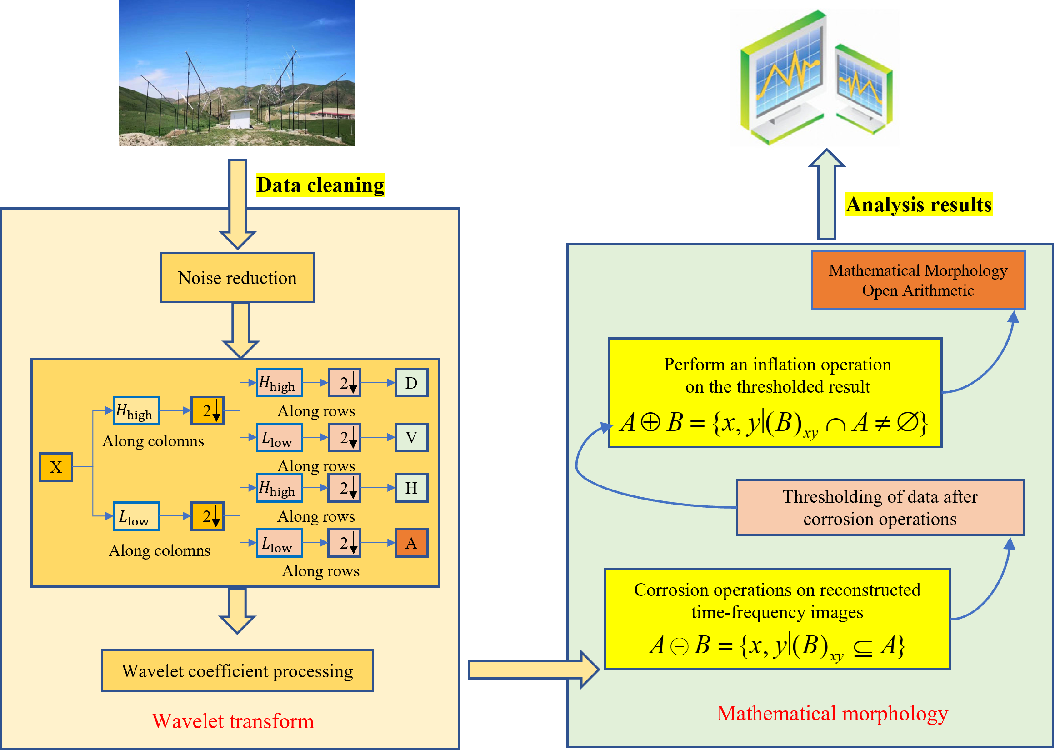}
   \caption{Noise reduction process using wavelet transform and mathematical morphology. }
   \label{Fig5}
   \end{figure}

The proposed RFI elimination method, based on the two-dimensional wavelet transform and mathematical morphology, effectively and quickly removes radio interference. Using the time-frequency series of the received data as input, the method first reduces the interference by subtracting the background noise. The data are then decomposed using the wavelet transform, and the RFI is further eliminated by processing the decomposed wavelet coefficients. Finally, mathematical morphology is applied to the reconstructed signal. Experimental results show that this method can effectively remove RFI while preserving signal quality. The specific algorithm principle is illustrated in (Fig.~\ref{Fig5}).
\subsection{Data Preprocessing}
Background subtraction is a classical and effective data processing technique used in radio observation. This method not only helps identify useful signals in dynamic spectra but also eliminates the influence of background noise. By subtracting the background noise, the analysis becomes cleaner and more accurate, allowing for a better study of the characteristics and effects of useful signals. The core of this method involves highlighting the signal changes by extracting a pre-recorded background signal and then subtracting it from the current observation data. This technique not only aids in accurately analyzing the target signal but also improves the flatness of the data, providing a reliable foundation for radio astronomy research. Its definition is shown in \autoref{eq4}:
\begin{equation}\label{eq4}
\begin{aligned}
  &P_{rec}=P_{\mathrm{sky}}+P_{\mathrm{sun}}\\
  &P_{rec}^{'}=P_{rec}-P_{\mathrm{sky}}
\end{aligned}  
\end{equation}
Where $P_\mathrm{sky}$ is the actual power of the background sky, $P_{rec}$ is the total power from the solar eruption, $P_\mathrm{sun}$ is the actual power of the solar eruption, and $P_{rec}^{'}$ is the output power after background subtraction.
\subsection{Two-dimensional Discrete Wavelet Transform}
\begin{figure}[h]
  \begin{minipage}[t]{0.495\linewidth}
  \centering
   \includegraphics[width=60mm]{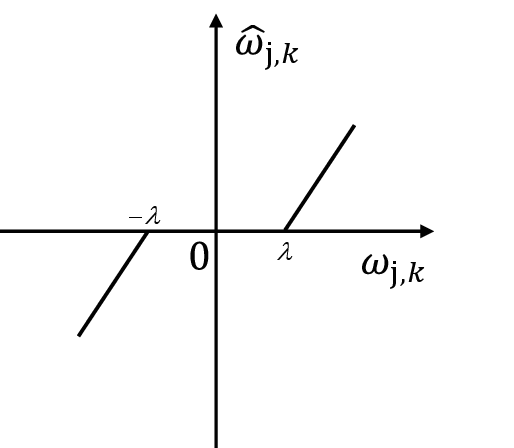}
	  \caption{\label{Fig6}{\small Hard threshold function diagram.} }
  \end{minipage}%
  \begin{minipage}[t]{0.495\textwidth}
  \centering
   \includegraphics[width=60mm]{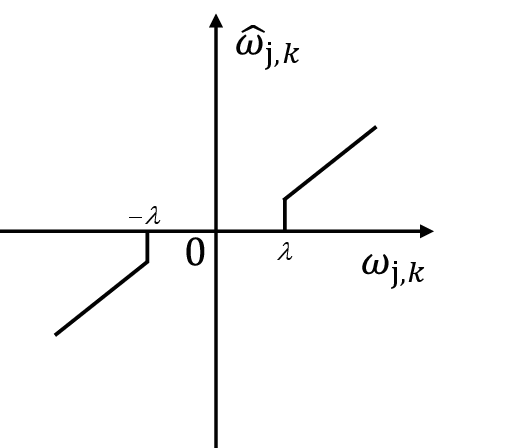}
	  \caption{\label{Fig7}{\small Soft threshold function diagram.}}
  \end{minipage}%
\end{figure}

\begin{figure}
   \centering
   \includegraphics[width=\textwidth, angle=0]{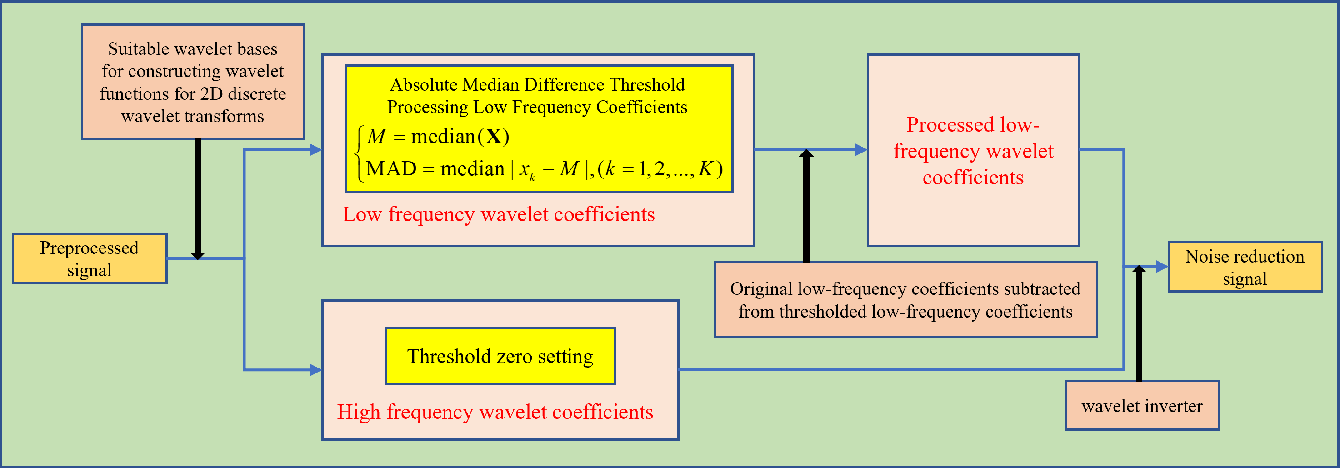}
   \caption{Flow chart of part of algorithm wavelet transform. }
   \label{Fig8}
   \end{figure}

Wavelet transform is an advanced mathematical signal processing technique primarily used to decompose complex signals or images into wavelet components at different frequencies. This enables a more accurate analysis of the signal’s characteristics and structure across multiple time scales (\citealt{32}). 
Compared to the traditional Fourier transform, the wavelet transform offers better time-domain localization and frequency-domain multi-resolution features. These advantages allow the wavelet transform to effectively capture sudden changes and local features in a signal, without needing to analyze the entire signal globally (\citealt{33}). Its flexibility and precision make the wavelet transform an essential tool for processing non-stationary signals and conducting multi-scale analysis.
Given the scale function $\varphi(x)$ and the wavelet function $\psi(x)$, a two-dimensional scale function and three two-dimensional wavelet functions can be combined, as shown in the mathematical definition of \autoref{eq5}:
\begin{equation}\label{eq5}
\begin{aligned}
&\varphi(x,y)=\varphi(x)\varphi(y) \\
&\psi^H(x,y)=\psi(x)\varphi(y) \\
&\psi^Y(x,y)=\varphi(x)\psi(y) \\
&\psi^D(x,y)=\psi(x)\psi(y)
\end{aligned}
\end{equation}
The input time-frequency series signal $f(x,y)$ can be represented as a linear combination of wavelet basis functions at multiple scales and positions. The expression for this representation is shown in \autoref{eq6}:
\begin{equation}\label{eq6}
f(x,y)=\frac1{\sqrt{MN}}\sum_m\sum_nW_\varphi(0,m,n)\varphi_{0,m,n}(x,y)+M
\end{equation}
The mathematical definition of $M$ is provided in \autoref{eq7}:
\begin{equation}\label{eq7}
M=\frac1{\sqrt{MN}}\sum_{j=0}^\infty\left\{M_1+M_2+M_3\right\}
\end{equation}
The mathematical definitions of $M_1$, $M_2$ and $M_3$ are provided in \autoref{eq8}:
\begin{equation}\label{eq8}
\begin{aligned}
&M_1=\sum_m\sum_\mathrm{n}W_\psi^H(j,m,n)\psi_{j,m,n}^H(x,y)\\
&M_2=\sum_m\sum_nW_\psi^V(j,m,n)\psi_{j,m,n}^V(x,y)\\
&M_3=\sum_m\sum_nW_\psi^D(j,m,n)\psi_{j,m,n}^D(x,y)    
\end{aligned}
\end{equation}
The mathematical definition of the approximation coefficient is given by the \autoref{eq9}:
\begin{equation}\label{eq9}
W_\varphi(0,m,n)=\frac1{\sqrt{MN}}\sum_{x=0}^{M-1}\sum_{y=0}^{N-1}f(x,y)\varphi_{0,m,n}(x,y)
\end{equation}
The mathematical definition of the detail coefficient is provided in \autoref{eq10}:
\begin{equation}\label{eq10}
W_\psi^i(j,m,n)=\frac1{\sqrt{MN}}\sum_{x=0}^{M-1}\sum_{y=0}^{N-1}f(x,y)\psi_{j,m,n}^i(x,y)\quad i=\{H,V,D\}
\end{equation}
Where $H$, $V$, and $D$ represent horizontal, vertical, and diagonal decomposition coefficients, respectively. $M$ and $N$ refer to the pixel dimensions $M*N$ of the input image, while $m$ and $n$ represent the translation parameters in the $x$ and $y$ directions, respectively.
$\varphi_{0,m,n}(x,y)$ and $\psi_{j,m,n}^i(x,y)$ are scale and wavelet functions at different scales and positions, respectively. Their mathematical definitions are given in \autoref{eq11} and \autoref{eq12}:
\begin{equation}\label{eq11}
\varphi_{0,m,n}(x,y)=2^{j/2}\varphi(2^jx-m,2^j-n)
\end{equation}
\begin{equation}\label{eq12}
\psi_{j,m,n}^i(x,y)=2^{j/2}\psi^i(2^jx-m,2^j-n)i=\{V,H,D\}
\end{equation}
Noise reduction methods based on wavelet transform can be categorized into three types (\citealt{19}). 
The first method involves wavelet thresholding, where smaller amplitude wavelet coefficients are treated as noise and suppressed by applying a threshold. 
The second method utilizes the local maximum principle of wavelet transform. This method exploits the differences in the propagation characteristics of signals and noise at various scales. It separates the signal from noise by identifying the mode maxima and reconstructing the wavelet coefficients using the residual mode maxima, which helps restore the original signal. 
The third method is based on the correlation principle of wavelet coefficients at different scales. This approach takes advantage of the strong correlation between coefficients across different scales to distinguish the signal from noise.\newline
In this section, the wavelet threshold denoising method is applied to process the measured radio signal (\citealt{34}).
Among the different techniques, hard thresholding is one of the simplest and most straightforward methods. The basic principle is to set the coefficients with absolute values less than or equal to the threshold to zero while retaining those with absolute values greater than the threshold. This method is particularly effective when the strong signal component is clearly distinguishable from the weak noise component, as it can effectively remove noise below the threshold while preserving the main signal information. However, when the difference between the signal and noise is not significant, using hard threshold denoising can lead to signal distortion and artifacts, often referred to as "edge artifacts." This occurs after the wavelet transform when the energy levels of the signal and noise are similar. In such cases, hard threshold denoising tends to over-truncate the coefficients, causing the processed signal to appear discontinuous and introducing a sawtooth effect. Its mathematical definition is shown in \autoref{eq13}:
\begin{equation}\label{eq13}
\hat{W}_{j,k}=\begin{cases}0,\mid W_{j,k}\mid\leqslant\lambda\\W_{j,k},\mid W_{j,k}\mid>\lambda&\end{cases}
\end{equation}
Soft Thresholding: The wavelet coefficients are processed by applying a threshold $\lambda$.
Specifically, the coefficients larger than the threshold are reduced by a certain amount, while those less than or equal to the threshold are set to zero. This method helps minimize the effect of noise while preserving the smoothness and key features of the signal. It is an effective technique for signal processing. The mathematical definition of this method is shown in \autoref{eq14}:
\begin{equation}\label{eq14}
\hat{W}_{j,k}=\begin{cases}0,\mid W_{j,k}\mid\leqslant\lambda\\\operatorname{sgn}(W_{j,k})(\mid W_{j,k}\mid-\lambda),\mid W_{j,k}\mid>\lambda\end{cases}
\end{equation}
Compared to hard threshold processing, the soft thresholding method not only smooths the signal but also introduces higher computational complexity. The advantage of this method is its ability to more accurately preserve the fine details of the signal, particularly the low-amplitude components. However, it may also cause some signal details to be lost or blurred during processing. The hard and soft threshold functions are shown in (Fig.~\ref{Fig6}) and (Fig.~\ref{Fig7}).\newline
In radio astronomy observation, the signals are often very weak and can easily be masked by background noise. However, wavelet analysis can significantly address this issue. After applying the wavelet transform, the energy of the effective signal is distributed across wavelet coefficients at different scales. This means that even if the signal is weak, it retains a relatively uniform energy distribution in the transformed data. This characteristic makes wavelet analysis an effective tool for detecting and extracting weak radio signals, enabling better detection and analysis of astronomical signals amidst complex background noise. In contrast, the energy of the RFI signal after wavelet transformation tends to be concentrated in a few key wavelet coefficients. By analyzing these significant coefficients, we can effectively detect and distinguish weak signals or interference hidden within the noise.\newline
Due to the diversity of RFI, it manifests itself differently in the wavelet coefficients. A sustained wideband RFI signal is typically more visible in the low-frequency coefficients of the wavelet transform. In the case of the two-dimensional discrete wavelet transform, signals with variability, such as broadband pulse signals in RFI, narrowband signals, or point-frequency instantaneous signals with abrupt changes, generally exhibit the following characteristics in the high-frequency coefficients: first, a straight line representing the frequency component of the signal, followed by a bright point that reflects the sudden or burst nature of the signal (\citealt{35}).
The reason for this phenomenon is that the mutational nature of these signals is more pronounced in the high-frequency coefficients, whereas in low-frequency coefficients, this effect may be less noticeable. This is because high-frequency coefficients typically correspond to rapid changes and detailed structures of the signal, making them more effective in capturing the instantaneous nature of sudden signal changes.\newline
In statistics, the Median Absolute Deviation (MAD) is a robust measure of the sample deviation for univariate numerical data. It can also represent the population parameter derived from the MAD estimate of the sample. As a result, the absolute median filter is applied to the coefficients after the wavelet transform. Additionally, MAD is more adaptable to outliers in the data set than the standard deviation (\citealt{36,37}). In the case of the standard deviation, the square of the distance from the data to the mean is used. Larger deviations receive higher weights, which means that outliers significantly influence the results. On the other hand, MAD is less sensitive to outliers, and a small number of them will not affect the outcome of the experiment. Its definition is given by \autoref{eq15}:
\begin{equation}\label{eq15}
\begin{cases}M=\text{median}(\mathbf{X})\\\text{MAD}=\text{median}\mid x_k-M\mid,(k=1,2,...,K)\end{cases}
\end{equation}
In the equation, $X(x_1,x_2,…,x_K)$ represents the given signal, where $K$ is the number of samples. First, the residual difference between the data and their median is calculated. The MAD is then defined as the median of the absolute values of these deviations.
\newline
(Fig.~\ref{Fig8}) shows the flow chart of the wavelet threshold denoising method used in this paper. A suitable wavelet basis is selected to construct the wavelet function. After transforming the time-frequency series data (with background noise removed) using the two-dimensional discrete wavelet transform, the wavelet coefficients of low frequency, horizontal high frequency, vertical high frequency, and diagonal high frequency are obtained. Threshold processing is then applied to eliminate certain outliers. Finally, the initial denoised signal is obtained through inverse wavelet transformation.
\subsection{Mathematical Morphology}
Mathematical morphology is a structured approach to image processing, grounded in set theory and algebraic theory, and mainly involves morphological operations. It can accurately identify and remove radio interference by analyzing the specific shape and structure of the signal. In addition, mathematical morphology offers efficient computational speed and real-time processing capabilities, allowing signal interference to be quickly addressed and ensuring the quality of communication. Furthermore, mathematical morphology can adjust its operational parameters and structural elements based on different signal characteristics and types of interference. This adaptability makes it highly effective in handling a variety of complex signal environments and interference types (\citealt{38,39,40,41}).\newline
Mathematical morphology involves two fundamental operations: erosion and dilation. The combination of erosion and dilation leads to the formation of the opening and closing operations  (\citealt{42}).
Dilation is the process by which background points in contact with an object are incorporated into the object under the constraint of a structural element B. As a result, the area of the object increases by the corresponding pixels. The mathematical definition of dilation is shown in \autoref{eq16}:
\begin{equation}\label{eq16}
A\oplus B=\{x,y|(B)_{xy}\cap A\neq\varnothing\}
\end{equation}
This equation represents the dilation of a target $A$ using a structuring element $B$, where the origin of $B$ is translated to the position of the image pixel $(x, y)$. If the intersection of $B$ and $A$ at the image element $(x, y)$ is not empty (that is, at least one of the image values corresponding to $A$ at the position of the element of 1 in $B$ is 1), the pixel $(x, y)$ in the output image is assigned the value 1. Otherwise, it is assigned 0. Thus, the dilation operation can be used to fill small holes in the target area and remove minor noise particles contained within the target.\newline
Erosion is the process of removing certain boundary points of an object using a structuring element. As a result, the area of the object is reduced by the corresponding number of points. Its definition is given in \autoref{eq17}:
\begin{equation}\label{eq17}
A\ominus B=\{x,y|(B)_{xy}\subseteq A\}
\end{equation}
It is important to define an origin for the structuring element $B$. When the origin of $B$ is shifted to the pixel $(x, y)$ of image $A$, if $B$ completely overlaps with the corresponding region in image $A$ (that is, all image values in $A$ corresponding to the positions of elements in $B$ that are equal to 1 are also 1), the output image at pixel $(x, y)$ is assigned a value of 1. Otherwise, the value is set to 0. Therefore, the erosion operation can effectively eliminate small, irrelevant objects. \newline
In mathematical morphology, the open and closed operations are image-processing techniques based on structural elements. The open operation is used to remove small objects while preserving the shape of the main structure. It works by first performing erosion and then dilation. The closed operation, on the other hand, performs dilation before erosion. It is mainly used to fill small holes within objects and smooth the boundaries of the objects (\citealt{43}).\newline
The selection of appropriate structuring elements is crucial for the effectiveness of morphological operations, as it directly impacts the precision of the operation and the extent of its influence. Common types of structuring elements include linear, square, and circular shapes, each with unique characteristics and suitable applications. The details are as follows:\newline
(1) Linear Structuring Elements: These are primarily used to process linear objects in images, allowing for control over the directionality and effect of morphological operations in specific cases. However, their effectiveness decreases when dealing with round or large-sized objects.\newline
(2) Square Structuring Elements: As simple and versatile elements, square structuring elements are suitable for most morphological operations and offer good adaptability. However, when working with non-square objects, they may not match the shape accurately, and complex image features might not be processed well.\newline
(3) Circular Structuring Elements: Circular elements are ideal for processing round or near-round targets. They provide higher accuracy in shape matching and filling effects. However, their computational complexity is high, particularly for large structuring elements. For non-circular targets, the results may be less satisfactory, and the operation’s effectiveness can be uncertain.
\section{Experimental Results and Analysis}
\subsection{Background Noise Analysis}
\begin{figure}[h]
  \begin{minipage}[t]{0.495\linewidth}
  \centering
   \includegraphics[width=60mm]{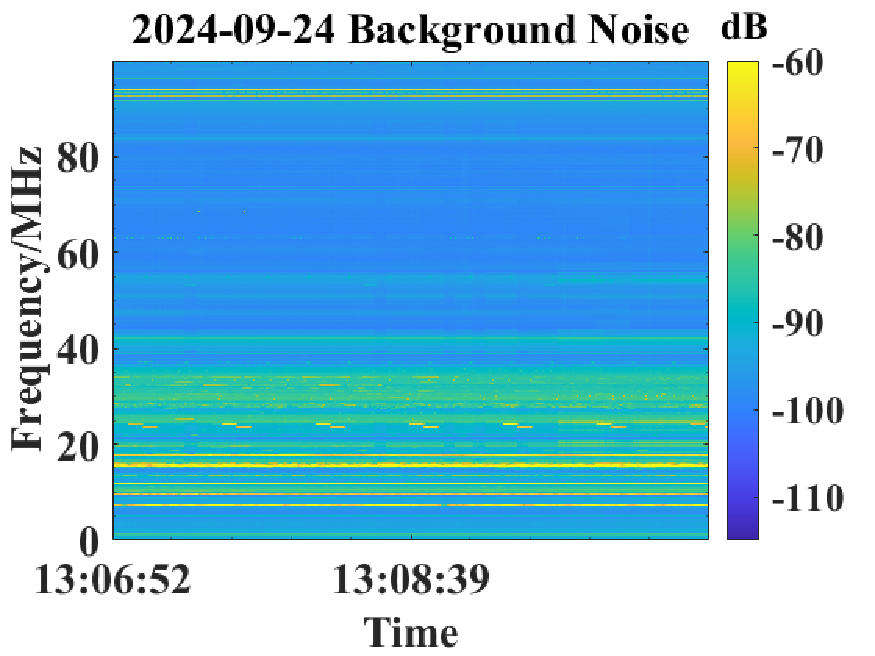}
	  \caption{\label{Fig9}{\small Background noise at the Qitai Station of the Xinjiang Observatory.} }
  \end{minipage}%
  \begin{minipage}[t]{0.495\textwidth}
  \centering
   \includegraphics[width=60mm]{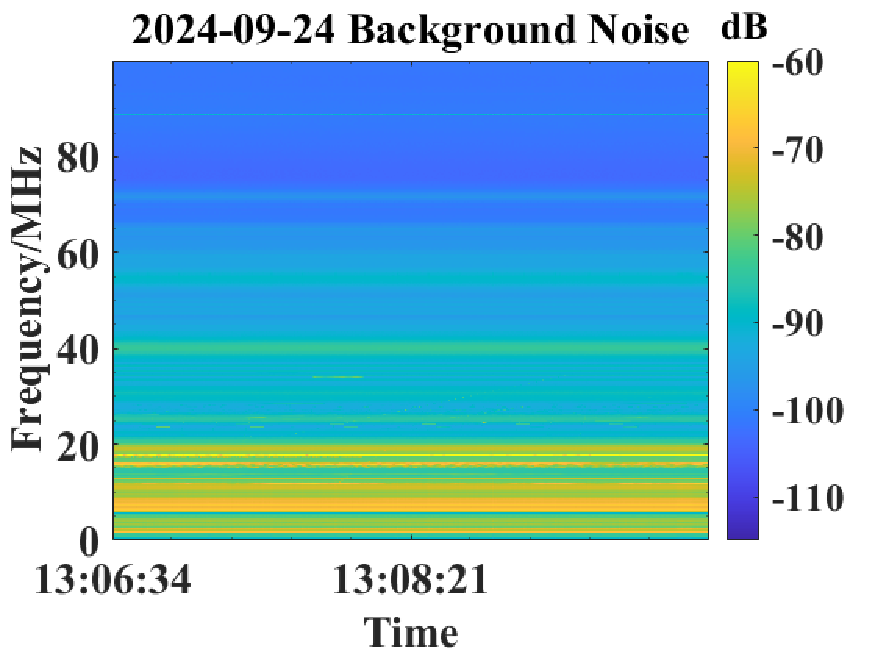}
	  \caption{\label{Fig10}{\small Background noise at the Yunnan Observatory.}}
  \end{minipage}%
\end{figure}

\begin{figure}[h]
  \begin{minipage}[t]{0.32\linewidth}
   \centering  
   \includegraphics[width=40mm]{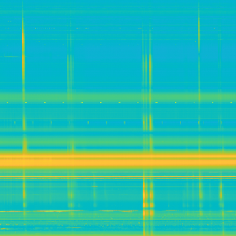}
   \caption{\label{Fig11}{Noise images 1.} }
  \end{minipage}%
  \begin{minipage}[t]{0.32\textwidth}
  \centering
   \includegraphics[width=40mm]{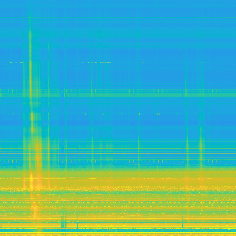}
	  \caption{\label{Fig12}{Noise images 2.}}
  \end{minipage}
    \begin{minipage}[t]{0.32\linewidth}
  \centering
   \includegraphics[width=40mm]{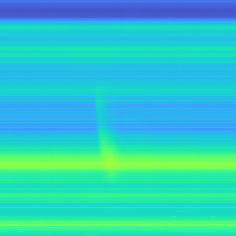}
	  \caption{\label{Fig13}{Noise images 3.} }
  \end{minipage}%
\end{figure}

\begin{figure}
   \centering
   \includegraphics[width=\textwidth, angle=0]{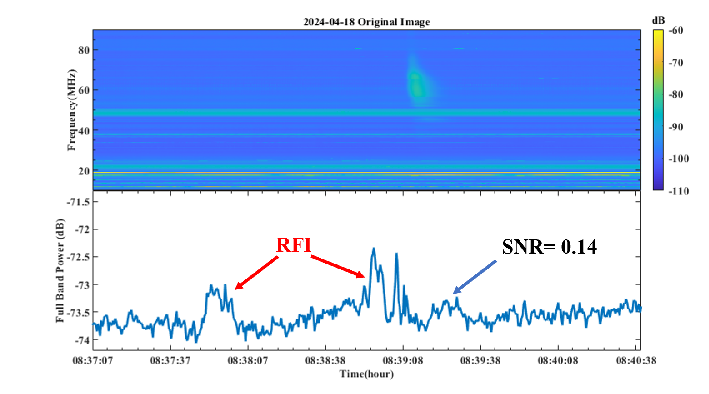}
   \caption{The top section shows the time-frequency spectrogram of the original image, while the bottom section presents the power change curve over time.}
   \label{Fig14}
   \end{figure}

\begin{figure}
   \centering
   \includegraphics[width=\textwidth, angle=0]{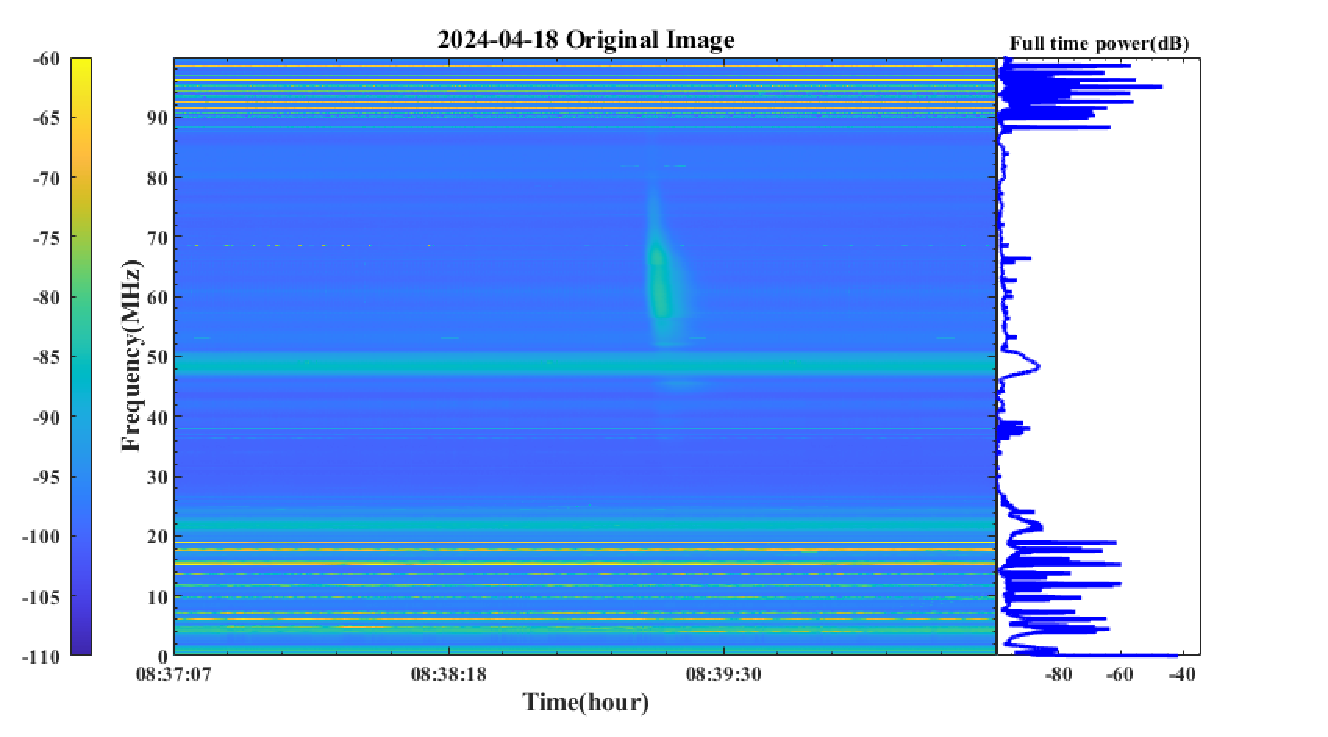}
   \caption{On the left is the time-frequency spectrogram of the original image, while the full-time power distribution curve is displayed on the right.}
   \label{Fig15}
   \end{figure}

\begin{figure}
   \centering
   \includegraphics[width=\textwidth, angle=0]{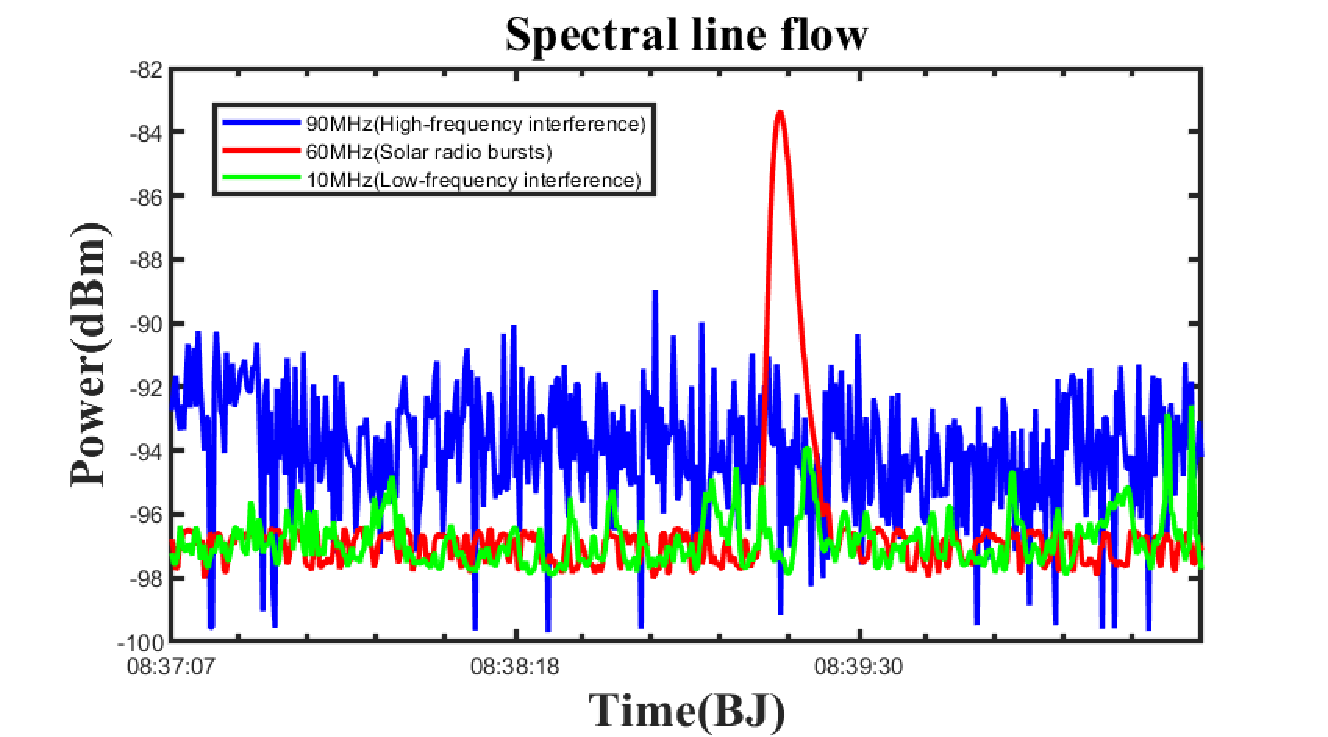}
   \caption{For the spectral line flow at selected frequency points in the original image, the blue curve represents the high-frequency NBI at 90 MHz, the red curve corresponds to the solar radio burst at 60 MHz, and the green curve indicates the low-frequency NBI at 10 MHz.}
   \label{Fig16}
   \end{figure}

In this paper, we have developed an eight-channel radio analyzer to monitor and analyze the quiet solar background noise at the Qitai Station of the Xinjiang Observatory and the Yunnan Observatory. The results show that RFI in the quiet solar background noise primarily originates from human activities, including shortwave AM broadcasting, single-sideband communication, FM broadcasting, data broadcasting, and television signals, as detailed in Table \ref{Table 1}. 
These interferences typically span a wide frequency range, disrupt the observed data spectrum, and degrade the quality of the data. They are commonly observed as strong horizontal fringes in the time-frequency graph. The shared characteristics of these interferences provide a crucial foundation for subsequent signal processing.\newline
\begin{table}[ht]
    \centering
    \caption{0-100 MHz Common artificial interference signals}
    \begin{tabular}{ccc}
        \toprule
        Wave band & Frequency (MHz) & Use  \\
        \midrule
        HF(SW) & 3.5-29.7 MHz & Shortwave AM broadcasting and single sideband communication \\
       VHF(FM) & 88-108 MHz & FM broadcasting and data broadcasting \\
        VHF & 48.5-92 MHz & Television and data broadcasting \\
        \bottomrule
    \end{tabular}
    \label{Table 1}
\end{table}
(Fig.~\ref{Fig9}) and (Fig.~\ref{Fig10}) present the time-frequency spectrogram of quiet solar background noise collected from both stations at 13:06 on September 24, 2024. Specifically, (Fig.~\ref{Fig9}) shows the observation data from the Qitai Station of the Xinjiang Observatory, while (Fig.~\ref{Fig10}) shows the data from the Yunnan Observatory.\newline
(Fig.~\ref{Fig11}), (Fig.~\ref{Fig12}), and (Fig.~\ref{Fig13}) show three groups of radio burst events that are affected by RFI. These NBIs cause discontinuities in the burst structure and obscure weak bursts, significantly affecting visual quality. This interference not only complicates the detection of solar radio bursts but also hampers the extraction of key information. In addition, the presence of noise reduces the detail and resolution of the image, resulting in poor image quality. This makes it more difficult to study the fine structure of solar radio bursts.
(Fig.~\ref{Fig14}) and (Fig.~\ref{Fig15}) display the time-frequency spectrograms of the Qitai Station at the Xinjiang Observatory, showing the frequency range of 0 to 100 MHz from 08:37:07 to 08:40:38 (Beijing time) on April 18, 2024. The time-frequency spectrograms illustrate the intensity of the solar radio signal as it varies over time and frequency. From the spectrograms, we observe a burst of instantaneous signal in the high-frequency band, approximately between 50 and 80 MHz, highlighted in the images. This may be related to a solar radio event. Although the signal time scale is short, the noise interference makes it difficult to discern its temporal characteristics. With a signal-to-noise ratio of only 0.14 dB, the interference signal significantly affects the overall power. Noise has severely impacted signal quality. In contrast, NBI appears as a smooth and continuous background with a more uniform energy distribution, but with localized intensifications at specific points, as indicated by the red arrow in (Fig.~\ref{Fig14})).\newline
From the perspective of the power distribution curve over the entire time period, the spectral characteristics of the solar radio signal are almost entirely obscured by noise. The power throughout the period is predominantly influenced by noise, making it difficult to identify the characteristics of the solar radio signal. The impact of NBI is particularly noticeable, mainly concentrated in the high-frequency range of 90 MHz to 100 MHz and the low-frequency range of 0 MHz to 20 MHz.\newline
(Fig.~\ref{Fig16})) illustrates the characteristics of high-frequency interference, low-frequency interference, and the signal strength of solar bursts across different frequency bands over time. The blue curve represents high-frequency signals, which are relatively strong and fluctuate over time. The green curve represents low-frequency signals, which maintain a relatively stable strength with minimal fluctuation. The stability of the low-frequency signal makes it easier to remove interference during the data pre-processing stage. However, both low-frequency and high-frequency interference exhibit sudden events, and these burst features are primarily reflected in the high-frequency coefficients after wavelet transformation, allowing for further elimination. Meanwhile, the red curve represents the solar burst signal, which shows that, although the interference effect is minor, the energy of the solar radio burst varies significantly. Despite this, small fluctuations are still observed in the signal during periods without bursts, and these fluctuations can affect the signal-to-noise ratio of the solar radio signal.
   
\subsection{Data Preprocessing Results and Analysis}
\begin{figure}[h]
  \begin{minipage}[t]{0.495\linewidth}
  \centering
   \includegraphics[width=60mm]{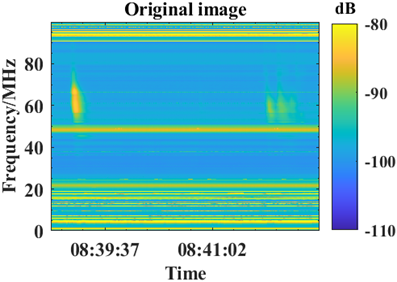}
	  \caption{\label{Fig17}{Original image.} }
  \end{minipage}%
  \begin{minipage}[t]{0.495\textwidth}
  \centering
   \includegraphics[width=60mm]{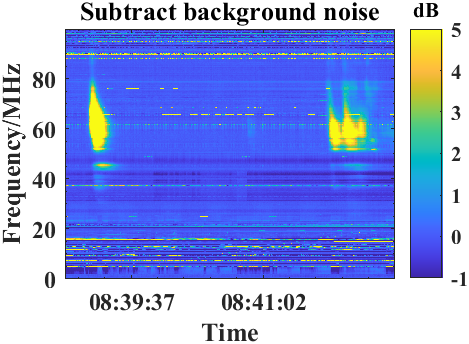}
	  \caption{\label{Fig18}{\small Subtracting background noise.}}
  \end{minipage}\\
    \begin{minipage}[t]{0.495\linewidth}
  \centering
   \includegraphics[width=60mm]{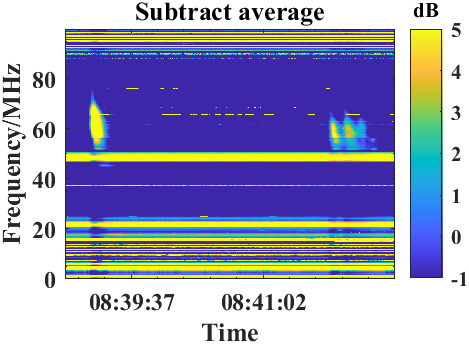}
	  \caption{\label{Fig19}{\small Subtracting channel means.} }
  \end{minipage}%
  \begin{minipage}[t]{0.495\textwidth}
  \centering
   \includegraphics[width=60mm]{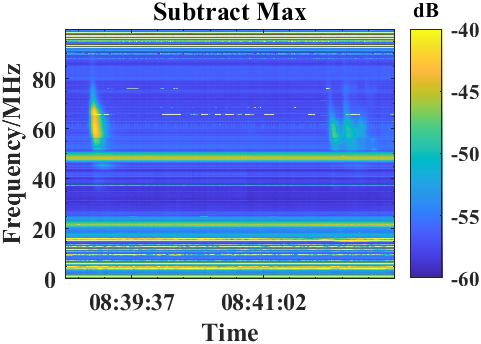}
	  \caption{\label{Fig20}{\small Subtracting channel maximum.}}
  \end{minipage}
\end{figure}

The solar radio signal is an abrupt random signal with significant transient characteristics, and its generation and disappearance usually occur in a very short period of time. Therefore, the pre-acquisition process of the background signal is relatively simple. However, in order to effectively avoid the interference of pre-acquisition signals that may contain radio signals, we refer to the real-time spectral data provided by the Learmonth Observatory in Australia (data source: https://www.sws.bom.gov.au/Solar/3/2/1). By analyzing the time periods in which no solar outbursts occurred in these data, we were able to accurately select the acquisition data from the observatory as the samples to be used in the preprocessing of the solar quiet background sky noise.\newline
(Fig.~\ref{Fig17}) shows the solar radio burst data received by the Qitai Station of the Xinjiang Observatory on April 18, 2024. In the original image, two Type III solar radio bursts are visible (\citealt{44}),along with significant NBI, some WBI, and bright spots, all of which severely affect subsequent data processing.
(Fig.~\ref{Fig18}) presents the radio spectrogram after subtracting the background noise.
By comparing Figure (Fig.~\ref{Fig17})) with (Fig.~\ref{Fig18}),it becomes clear that, after background noise subtraction, the Type III solar burst near 60 MHz is more clearly displayed. Compared to the channel effect removal methods of subtracting the mean value of the corresponding channel in (Fig.~\ref{Fig19}) and subtracting the maximum value of the corresponding channel in (Fig.~\ref{Fig20}), (Fig.~\ref{Fig18}) exhibits fewer RFI and clearer burst information. These results fully demonstrate the feasibility and effectiveness of the proposed method.
\subsection{Experimental Results and Analysis of Wavelet Transform}
\begin{figure}
   \centering
   \includegraphics[width=\textwidth, angle=0]{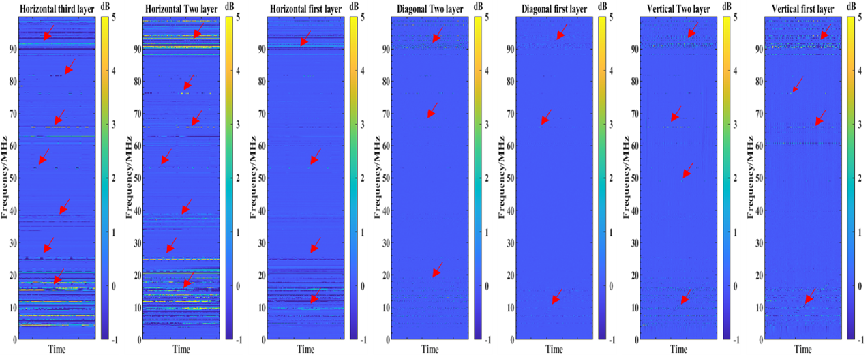}
   \caption{ Remapping of high-frequency coefficients. }
   \label{Fig21}
   \end{figure}

\begin{figure}
   \centering
   \includegraphics[width=\textwidth, angle=0]{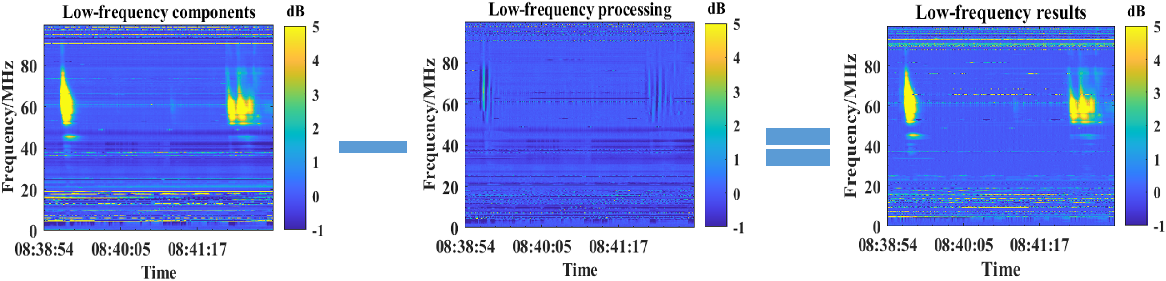}
   \caption{Processing of low-frequency coefficients. }
   \label{Fig22}
   \end{figure}

\begin{figure}
   \centering
   \includegraphics[width=\textwidth, angle=0]{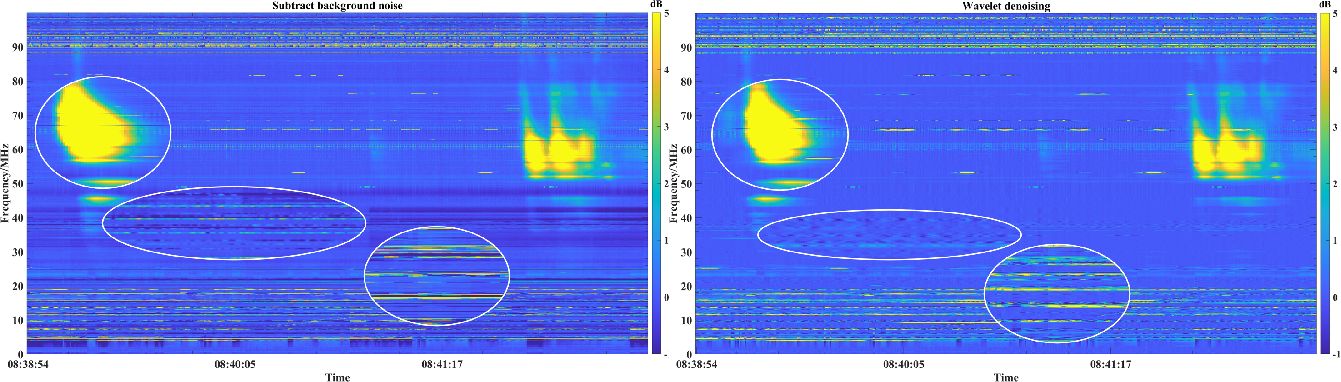}
   \caption{ Comparison of RFI removal and background noise elimination through wavelet transform. }
   \label{Fig23}
   \end{figure}

\begin{figure}[h]
  \begin{minipage}[t]{0.495\linewidth}
  \centering
   \includegraphics[width=60mm]{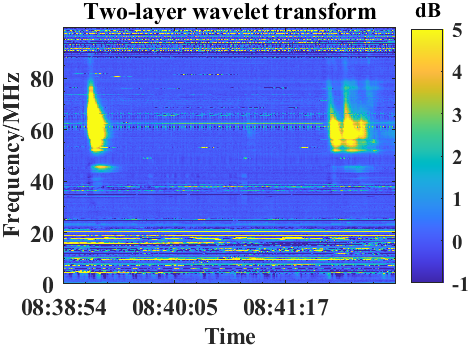}
	  \caption{\label{Fig24}{\small Two-layer wavelet decomposition.} }
  \end{minipage}%
  \begin{minipage}[t]{0.495\textwidth}
  \centering
   \includegraphics[width=60mm]{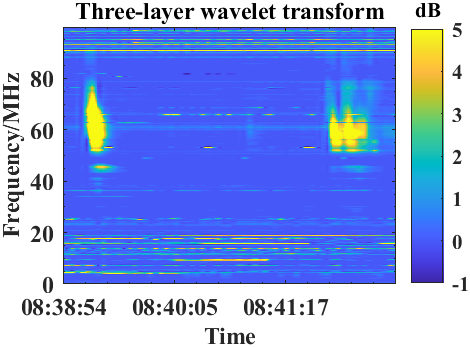}
	  \caption{\label{Fig25}{\small Three-layer wavelet decomposition.}}
  \end{minipage}\\
    \begin{minipage}[t]{0.495\linewidth}
  \centering
   \includegraphics[width=60mm]{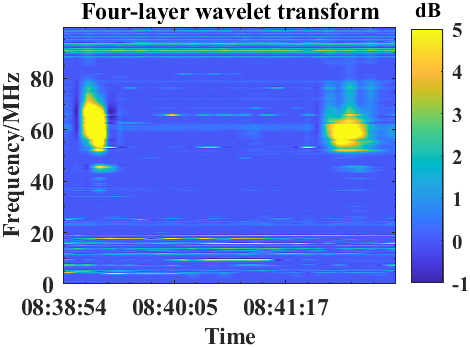}
	  \caption{\label{Fig26}{\small Four-layer wavelet decomposition.} }
  \end{minipage}%
  \begin{minipage}[t]{0.495\textwidth}
  \centering
   \includegraphics[width=60mm]{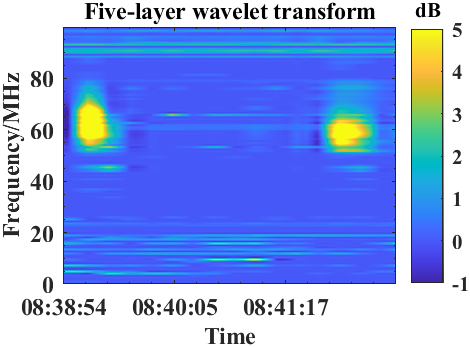}
	  \caption{\label{Fig27}{\small Five-layer wavelet decomposition.}}
  \end{minipage}
\end{figure}

Firstly, the time-frequency spectrogram with subtracted background noise is decomposed using a three-layer two-dimensional discrete wavelet transform, obtaining both low-frequency and high-frequency components. In this example, the high-frequency components after the three-layer decomposition of the reconstructed part, including the horizontal, vertical, and diagonal components, are shown in (Fig.~\ref{Fig21}).
These high-frequency components do not contain broadband solar radio signals but only exhibit mutational NBI and isolated bright spot interference, as indicated by the red arrow in (Fig.~\ref{Fig21})).\newline
The high-frequency coefficients are set to zero using a threshold value to eliminate high-frequency interference. The low-frequency coefficients are then processed using the MAD threshold input; coefficients above the threshold are set to zero, and the difference between the original coefficients and the thresholded coefficients is calculated to preserve more useful signals. The processing steps are shown in (Fig.~\ref{Fig22}). Using the processed wavelet coefficients, the original spectral data is then reconstructed through an inverse wavelet transform. The inverse wavelet transform is the process of transforming the thresholded wavelet coefficients back to the time domain, resulting in the reconstructed image without RFI.\newline
(Fig.~\ref{Fig23}) shows the time-frequency spectrogram after the removal of background noise and reconstruction of the wavelet transform. The left image displays the result of background noise removal, while the right image shows the result after RFI removal using a wavelet transform. Comparing the two images, it can be seen that after the wavelet transform, most NBI and bright spot interferences are eliminated without losing the key information in the image. Additionally, continuous NBI is converted into isolated points, which facilitates the subsequent mathematical morphology processing.\newline
In wavelet transform, increasing the number of decomposition layers can more effectively separate the noise and signal features, which aids in their removal. However, as the number of decomposition layers increases, the reconstructed spectrum signal may experience significant distortion, negatively impacting the final noise reduction effect. Additionally, the wavelet transform typically performs further decomposition on the low-frequency components of the spectrum data, while the high-frequency components are not subdivided. Therefore, increasing the number of decomposition layers may lead to a loss of spectral signal details, whereas reducing the number of layers may not effectively eliminate high-frequency noise.
(Fig.~\ref{Fig24}) shows the results after a two-layer wavelet decomposition and noise reduction. The radio spectrum signal does not exhibit significant distortion, but the RFI remains noticeable. 
(Fig.~\ref{Fig25})presents the results after a three-layer wavelet decomposition and noise reduction. The spectrum signal shows no obvious distortion, and most of the RFI is eliminated. 
In (Fig.~\ref{Fig26}) and (Fig.~\ref{Fig27}), four- and five-layer wavelet decompositions cause noticeable spectrum signal distortion. Therefore, this study chooses the three-layer wavelet decomposition as the optimal choice for radio spectrum signal processing.
\subsection{Experimental Results and Analysis of Mathematical Morphology Processing}
\begin{figure}[h]
  \begin{minipage}[t]{0.495\linewidth}
  \centering
   \includegraphics[width=60mm]{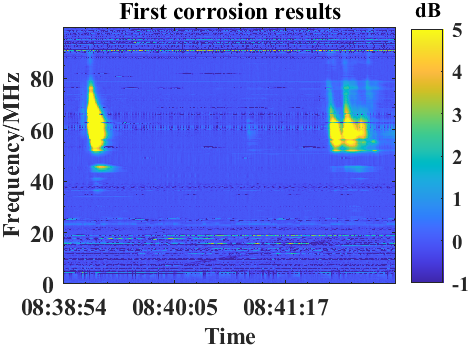}
	  \caption{\label{Fig28}{\small The first corrosion results.} }
  \end{minipage}%
  \begin{minipage}[t]{0.495\textwidth}
  \centering
   \includegraphics[width=60mm]{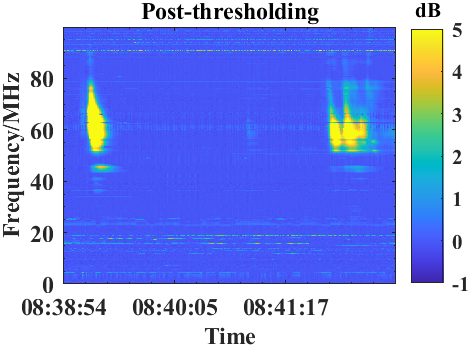}
	  \caption{\label{Fig29}{\small First corrosion threshold treatment.}}
  \end{minipage}%
\end{figure}

\begin{figure}[h]
  \begin{minipage}[t]{0.495\linewidth}
  \centering
   \includegraphics[width=60mm]{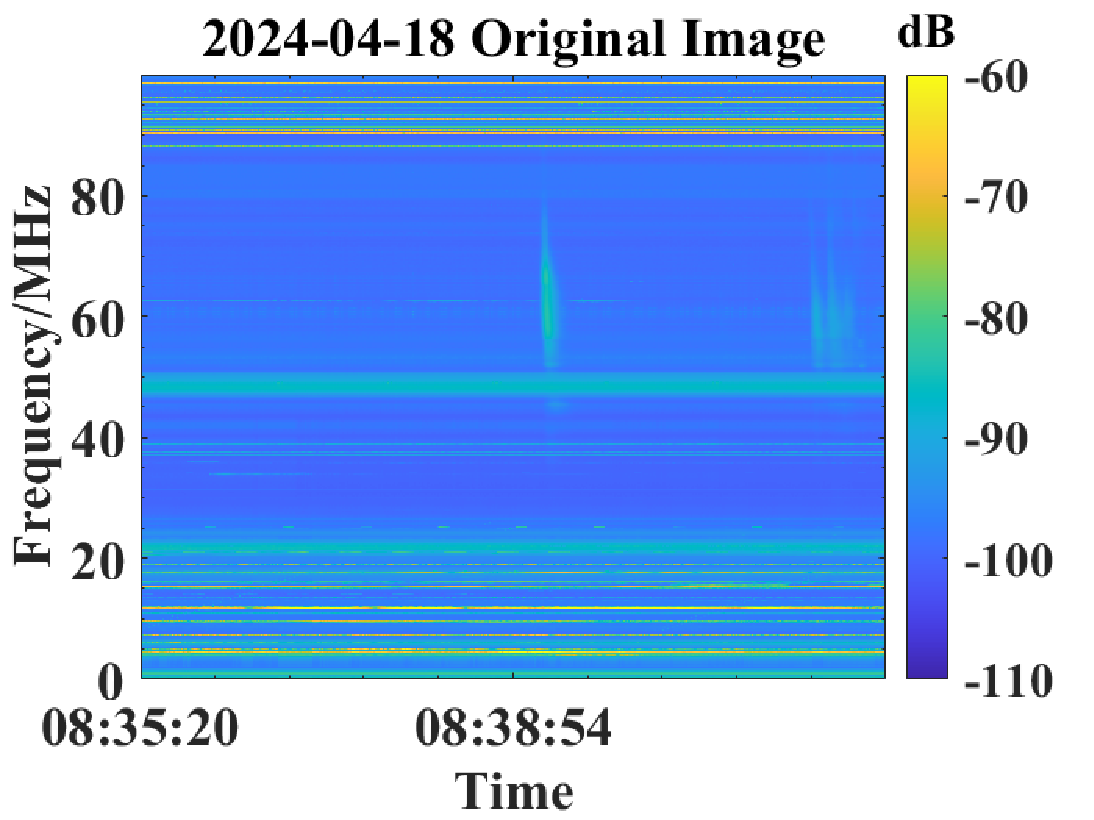}
	  \caption{\label{Fig30}{\small Original image.} }
  \end{minipage}%
  \begin{minipage}[t]{0.495\textwidth}
  \centering
   \includegraphics[width=60mm]{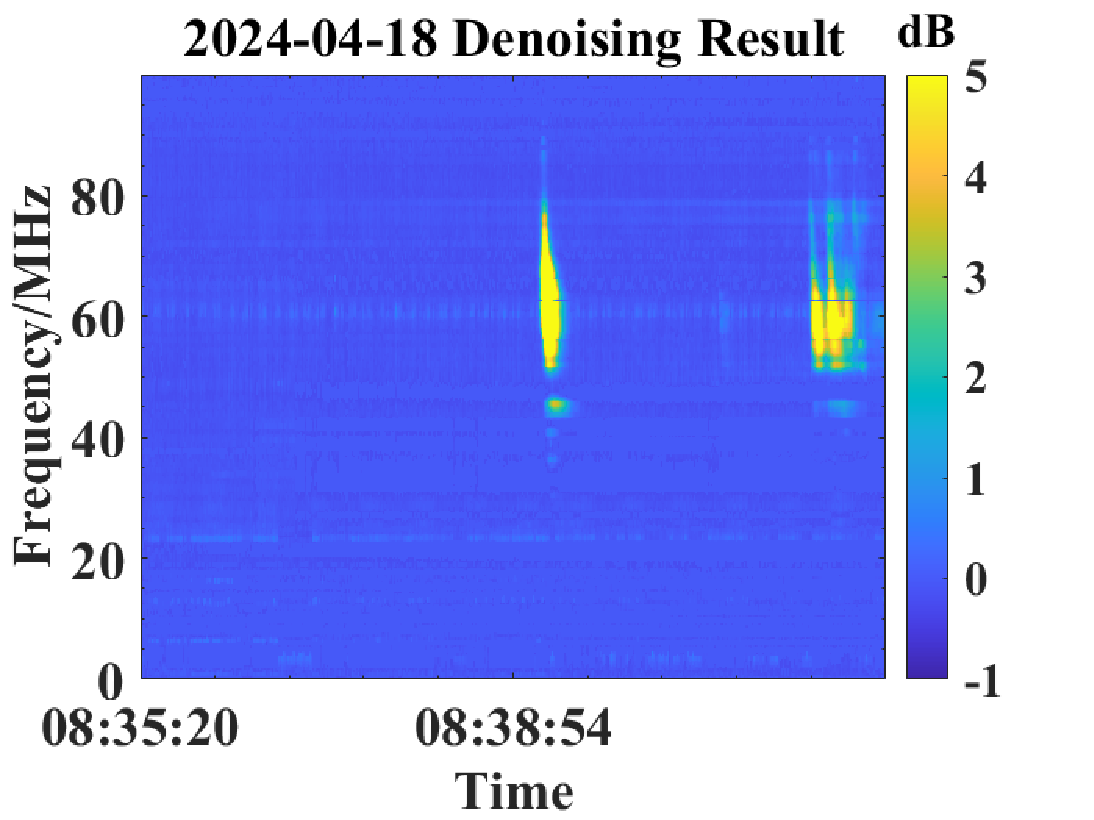}
	  \caption{\label{Fig31}{\small Denoising result.}}
  \end{minipage}%
\end{figure}

\begin{figure}[h]
  \begin{minipage}[t]{0.495\linewidth}
  \centering
   \includegraphics[width=60mm]{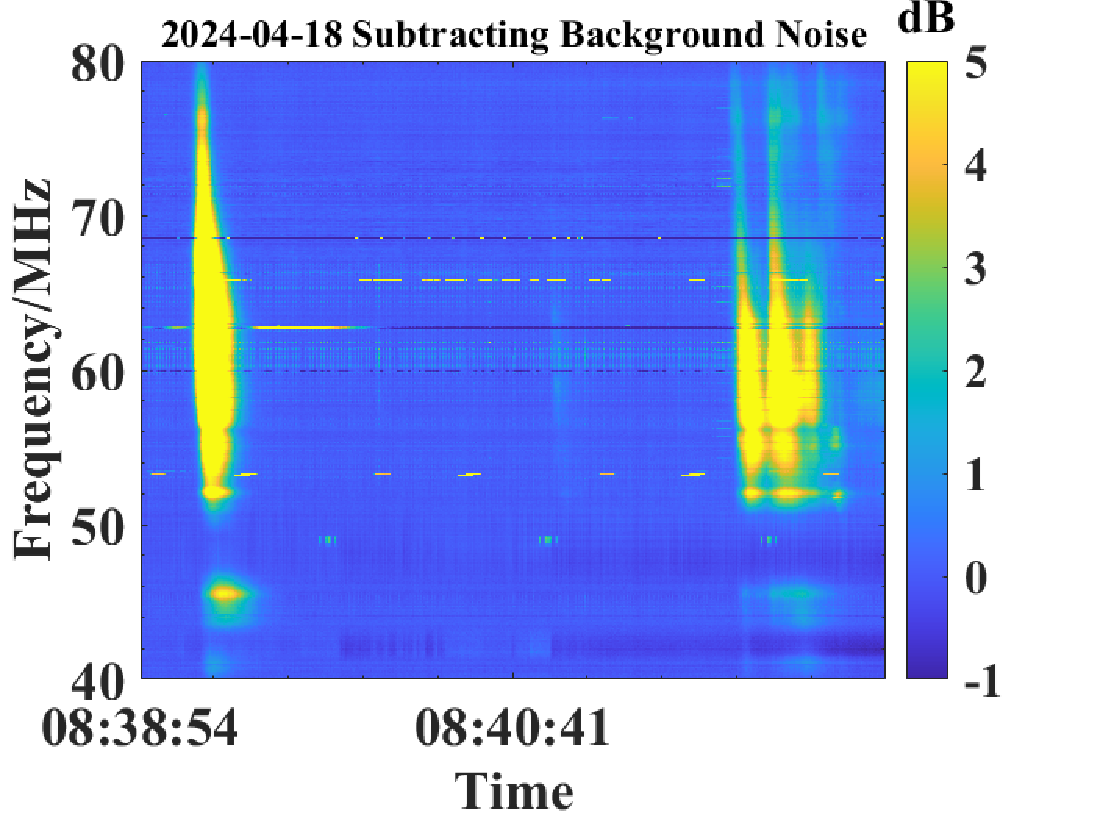}
	  \caption{\label{Fig32}{\small Subtracting background noise.} }
  \end{minipage}%
  \begin{minipage}[t]{0.495\textwidth}
  \centering
   \includegraphics[width=60mm]{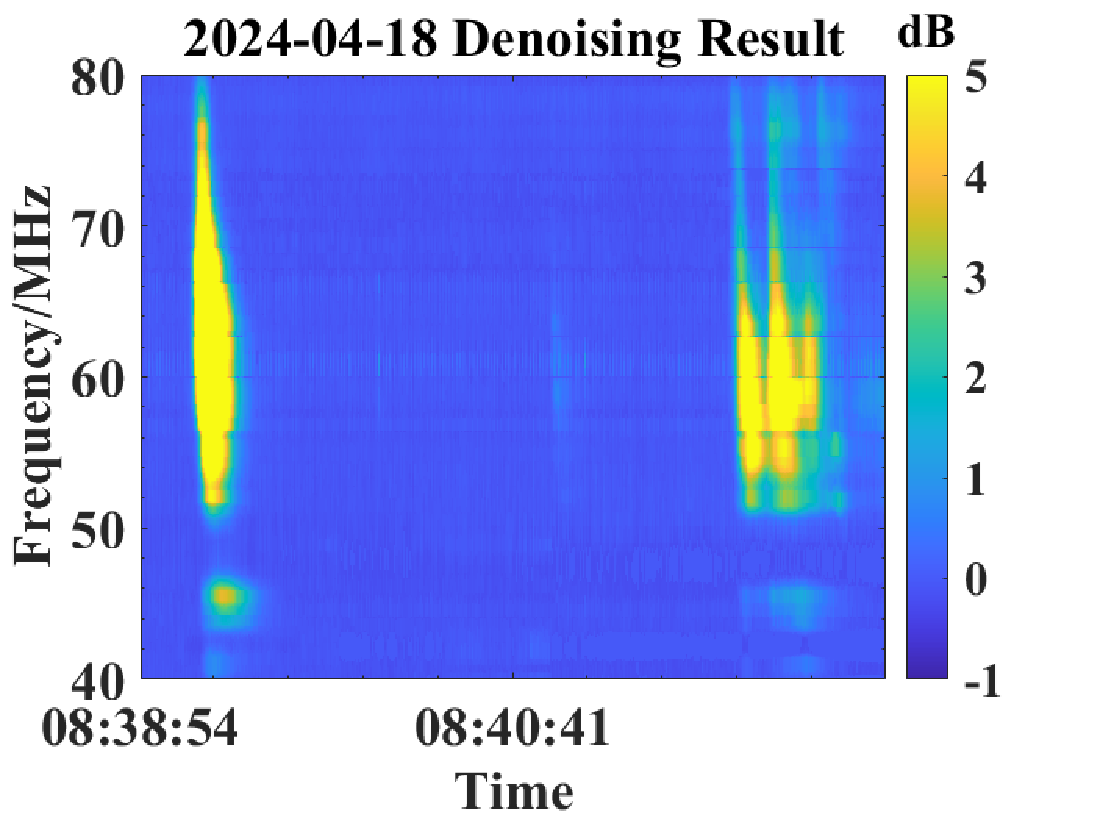}
	  \caption{\label{Fig33}{\small Linear structure.}}
  \end{minipage}\\
    \begin{minipage}[t]{0.495\linewidth}
  \centering
   \includegraphics[width=60mm]{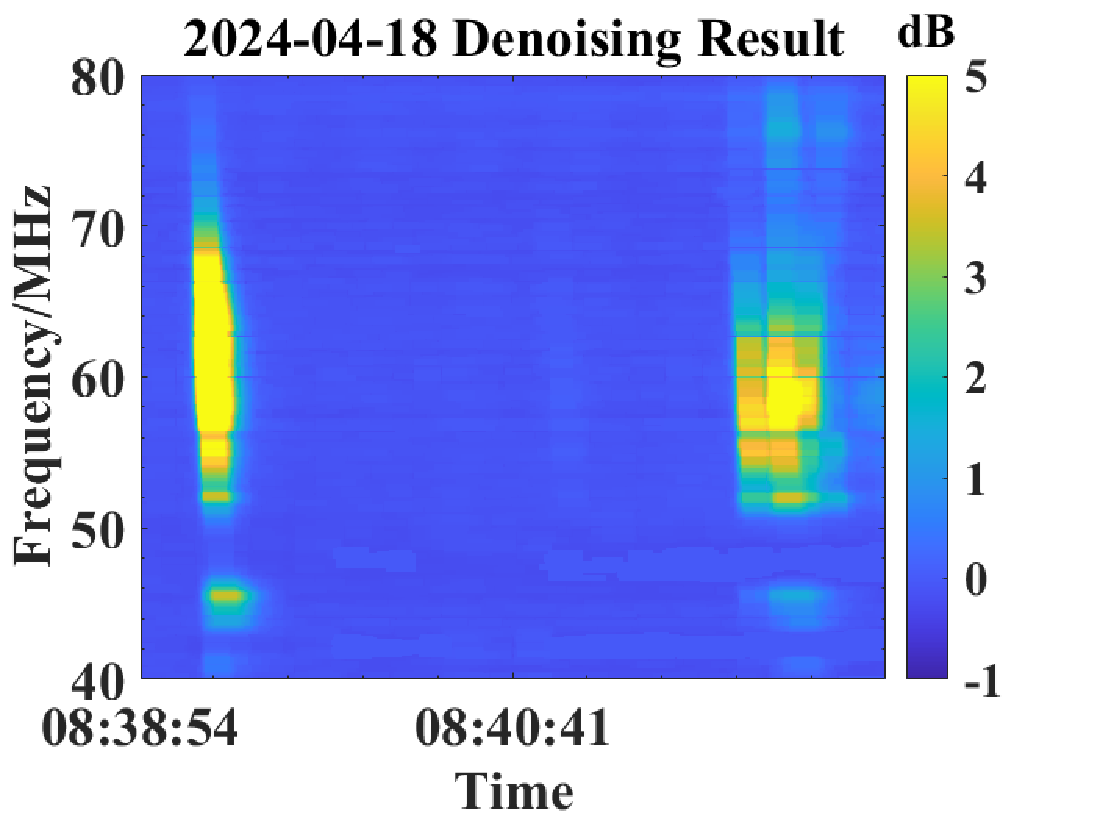}
	  \caption{\label{Fig34}{\small Square structure.} }
  \end{minipage}%
  \begin{minipage}[t]{0.495\textwidth}
  \centering
   \includegraphics[width=60mm]{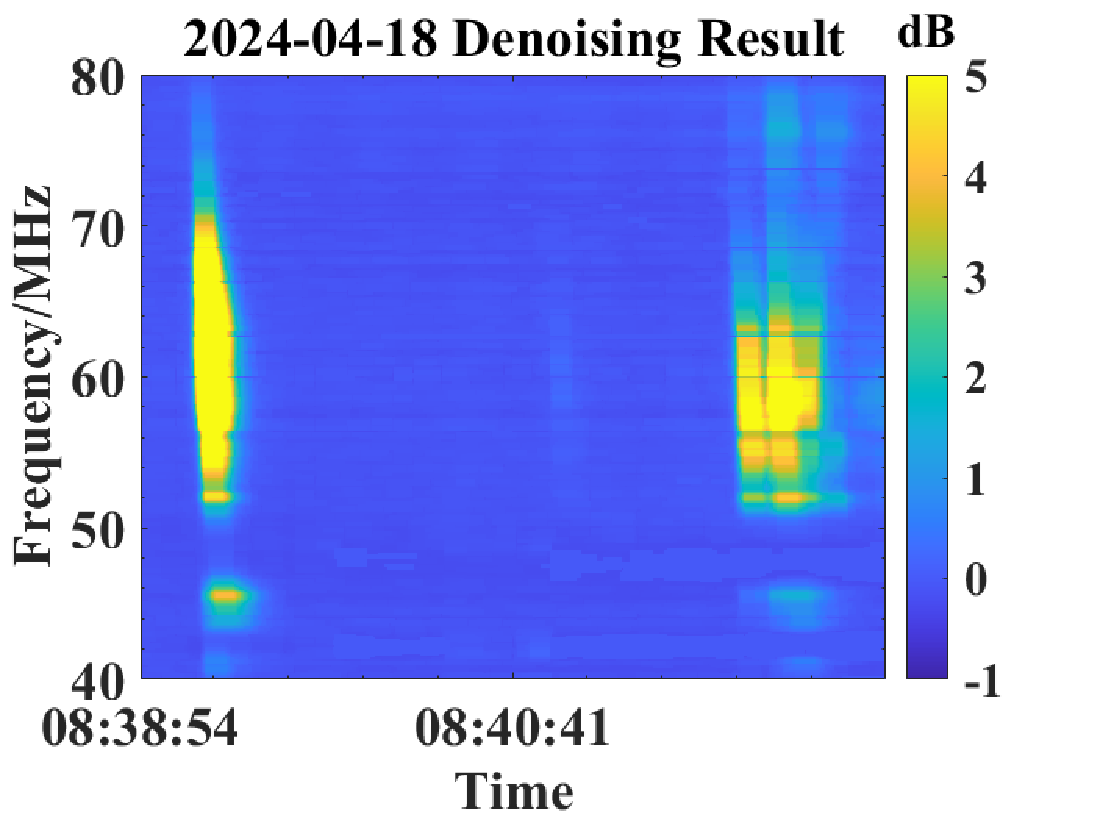}
	  \caption{\label{Fig35}{\small Circular structure.}}
  \end{minipage}
\end{figure}

In this paper, mathematical morphology operations are applied based on wavelet transform processing. First, a mathematical morphological erosion operation is performed on the data. By sliding the structuring element across the image, the pixel under the element is set to the minimum value within the covered area. This effectively eliminates small targets in the image and smooths the edges of the object, as shown in (Fig.~\ref{Fig28}). 
This operation removes most of the NBI from the time-frequency spectrogram. However, artifacts are still visible after interference is eliminated, and a small amount of interference remains. To address this, threshold processing is applied to the image after the corrosion operation to eliminate the artifacts. The result after elimination is shown in (Fig.~\ref{Fig29}).\newline
Although the RFI is eliminated after the erosion operation, the spectrum data image suffers from the separation and loss of some data information. To address this issue, this paper performs a dilation operation on the image processed by the initial erosion and threshold steps. This operation increases the size of the target in the image and fills the gaps between the objects. 
After erosion and dilation, some RFI remains, so the image is further processed using mathematical morphology operations. The opening operation is applied to remove small RFI points while preserving the size of the desired objects. The final data processing results are shown in (Fig.~\ref{Fig30}) and (Fig.~\ref{Fig31}).\newline
\newline
In mathematical morphology, selecting the appropriate size for structural elements is a delicate balance between maintaining operational efficiency and preventing overprocessing or loss of detail. In this paper, we focus on narrow-band interference in the solar radio spectrum, which is typically characterized by long, straight lines in time-frequency spectrograms. After applying the two-dimensional discrete wavelet transform, some isolated linear interference remains in the solar radio spectrum. Given this, the linear structural element in mathematical morphology proves to be particularly effective in addressing such issues, as it can efficiently remove these interferences.\newline
Therefore, this paper adopts the empirical formula for selecting structural elements:
\begin{equation}\label{eq18}
L=\alpha \cdot \lambda
\end{equation}
Where $L$ is the length of the structural element,  $\lambda$ is the interference length of the target signal, and $\alpha$ is the  empirical coefficient, with a value range of  0.8 to 1.2.\newline
Based on \autoref{eq18}, linear structural elements of various sizes were tested in comparative experiments. The 10×10 linear structural element was ultimately selected to effectively suppress interference while preserving the features of the original signal as much as possible.\newline
To validate the best denoising performance, this paper tests structural elements of various shapes, including linear, rectangular, and circular.
(Fig.~\ref{Fig33}), (Fig.~\ref{Fig34}), and (Fig.~\ref{Fig35}) present the final denoising results. While rectangular and circular structural elements can effectively remove RFI, they also lead to some loss of the radio burst spectrum signal data.
In comparison, (Fig.~\ref{Fig32}) shows that the linear structural element not only preserves the quality of the spectrum data but also effectively eliminates RFI.\newline
\subsection{Performance Analysis}
\begin{figure}[h]
    \begin{minipage}[t]{0.495\linewidth}
  \centering
   \includegraphics[width=60mm]{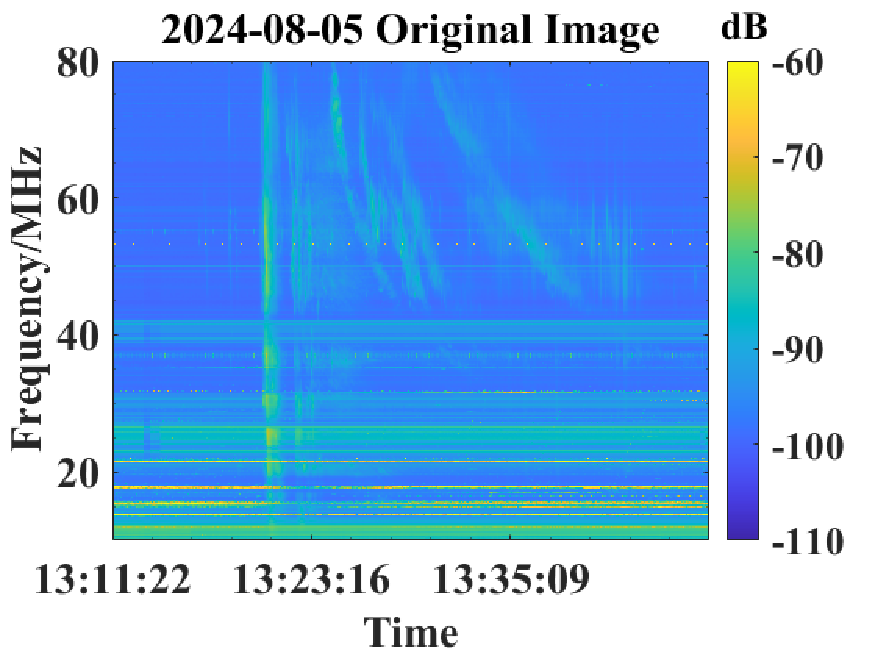}
	  \caption{\label{Fig36}{\small Original image of the burst cluster.} }
  \end{minipage}%
  \begin{minipage}[t]{0.495\textwidth}
  \centering
   \includegraphics[width=60mm]{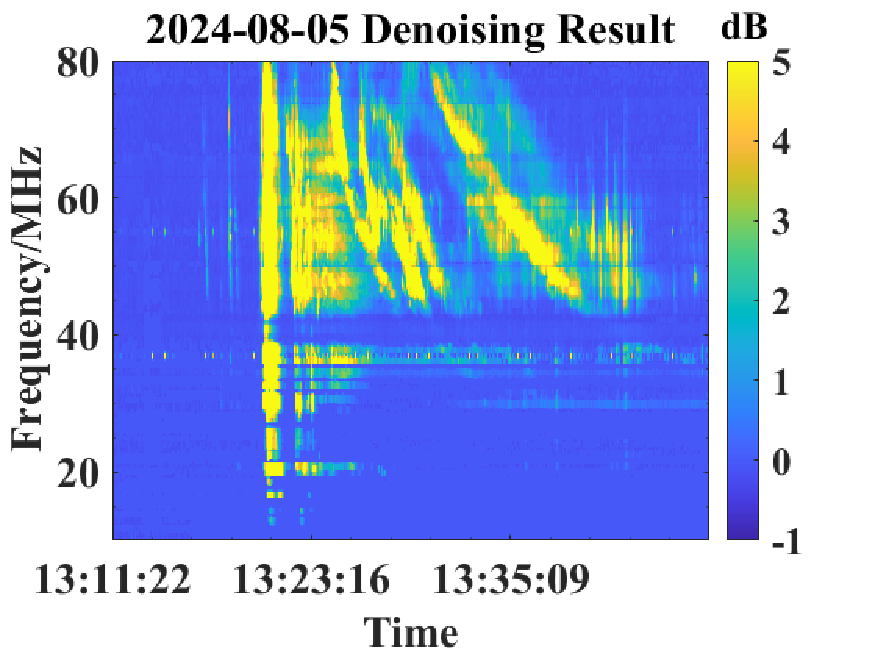}
	  \caption{\label{Fig37}{\small Denoising result of the burst cluster.}}
  \end{minipage}
  \begin{minipage}[t]{0.495\linewidth}
  \centering
   \includegraphics[width=60mm]{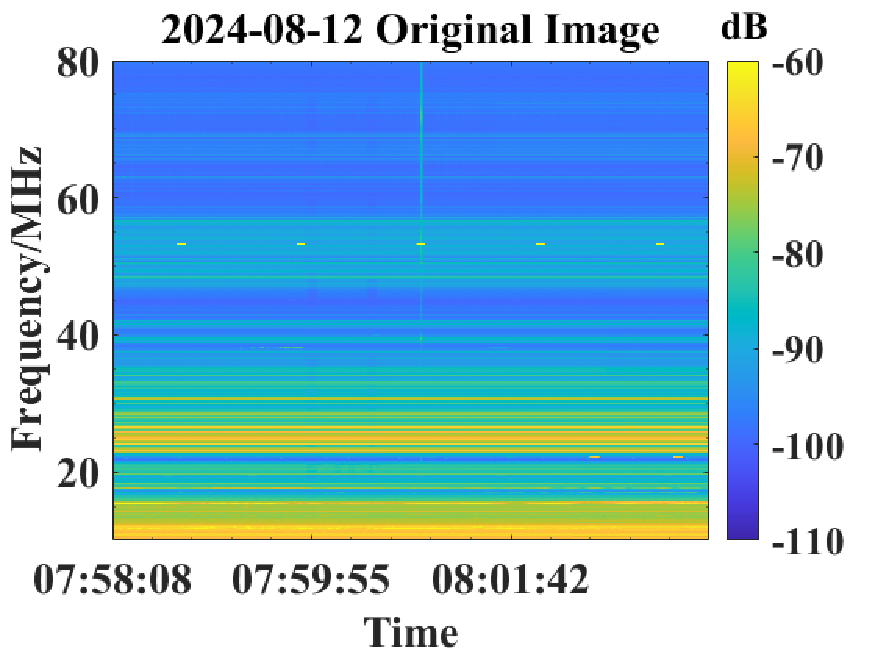}
	  \caption{\label{Fig38}{\small Original image of Yunnan Observatory.} }
  \end{minipage}%
  \begin{minipage}[t]{0.495\textwidth}
  \centering
   \includegraphics[width=60mm]{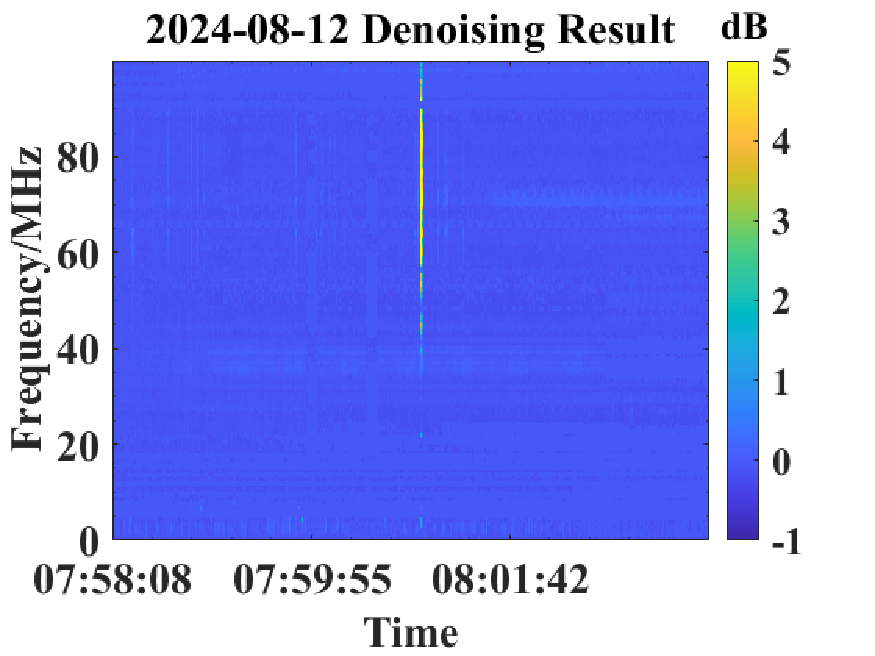}
	  \caption{\label{Fig39}{\small Denoising result of Yunnan Observatory.}}
  \end{minipage}\\
    \begin{minipage}[t]{0.495\linewidth}
  \centering
   \includegraphics[width=60mm]{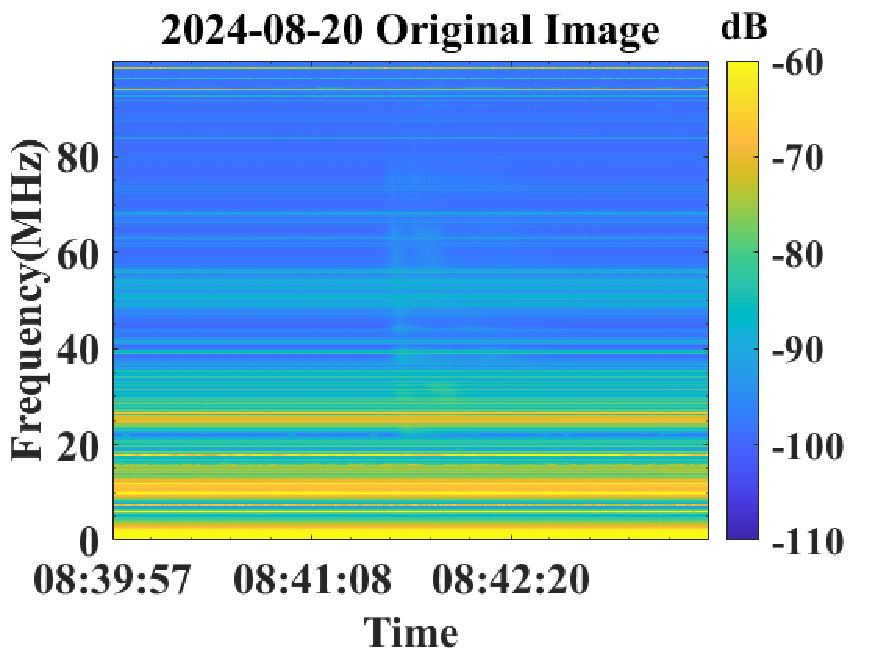}
	  \caption{\label{Fig40}{\small Original image of  Mianyang Liangdan City.} }
  \end{minipage}%
  \begin{minipage}[t]{0.495\textwidth}
  \centering
   \includegraphics[width=60mm]{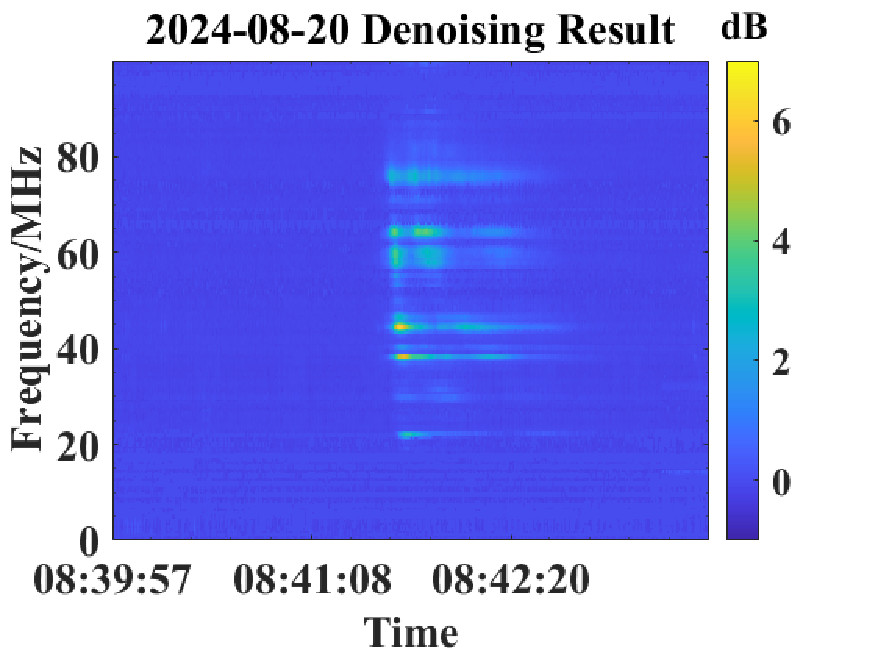}
	  \caption{\label{Fig41}{\small Denoising result of Mianyang Liangdan City.}}
  \end{minipage}\\
\end{figure}

\begin{figure}[h]
  \begin{minipage}[t]{0.495\linewidth}
  \centering
   \includegraphics[width=60mm]{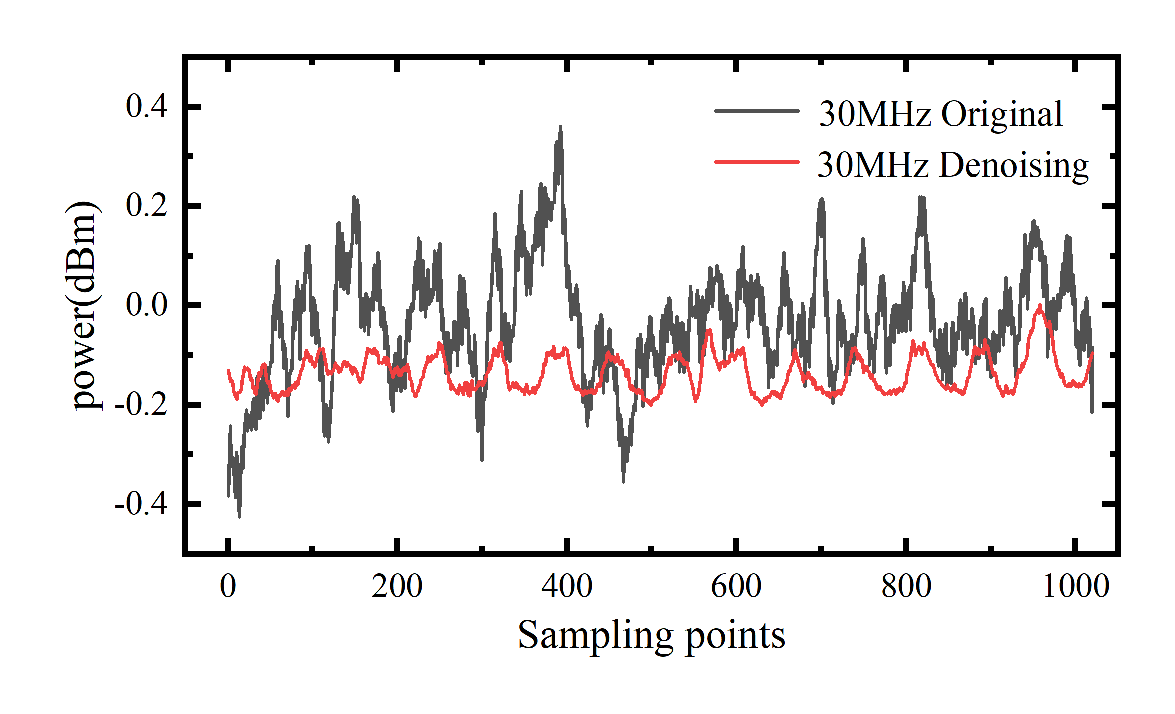}
	  \caption{\label{Fig42}{\small 30MHz spectral line flow.} }
  \end{minipage}%
  \begin{minipage}[t]{0.495\textwidth}
  \centering
   \includegraphics[width=60mm]{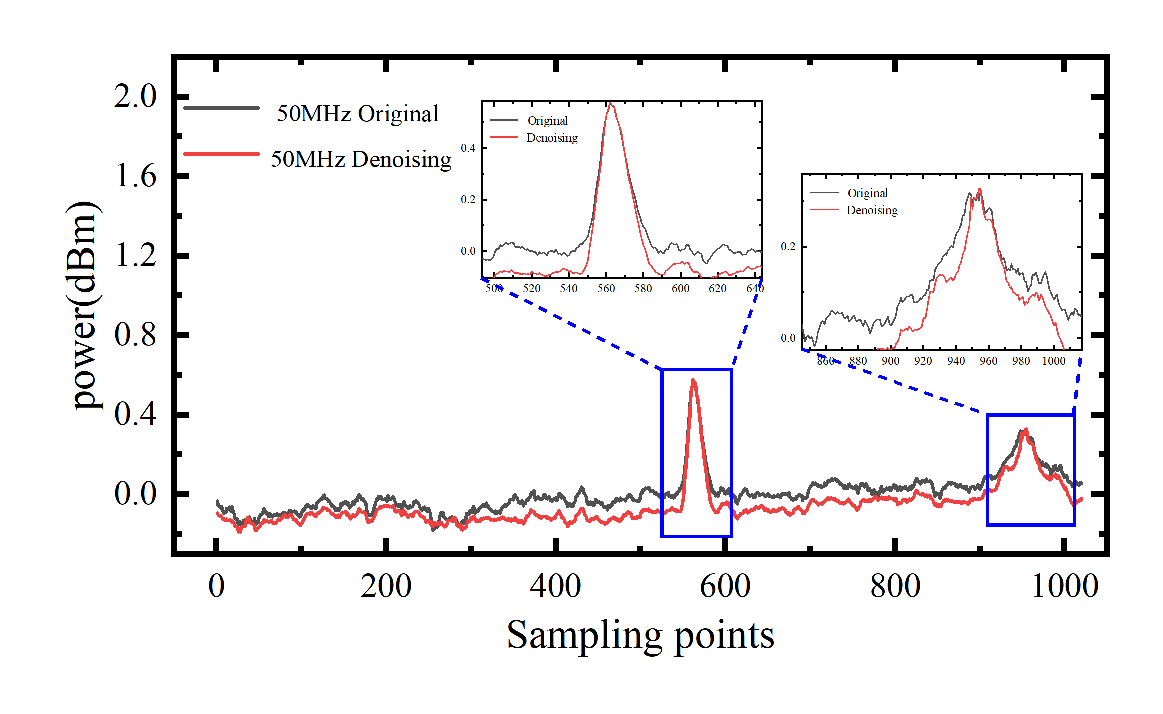}
	  \caption{\label{Fig43}{\small 50MHz spectral line flow.}}
  \end{minipage}\\
    \begin{minipage}[t]{0.495\linewidth}
  \centering
   \includegraphics[width=60mm]{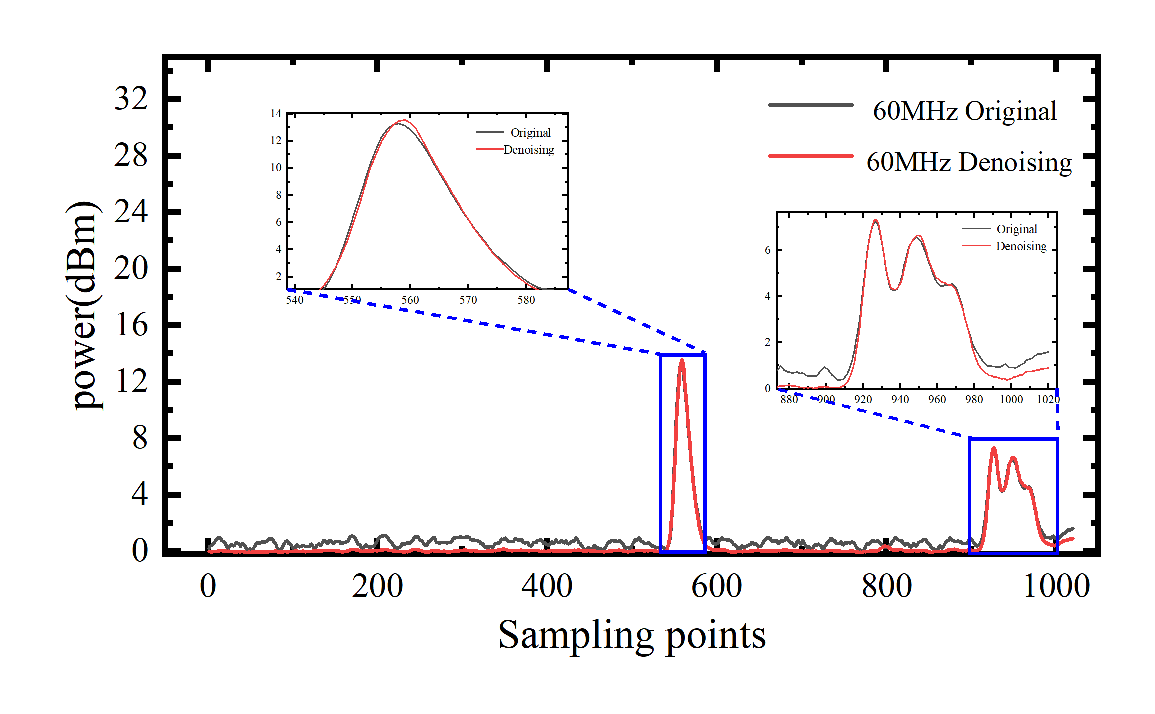}
	  \caption{\label{Fig44}{\small 60MHz spectral line flow.} }
  \end{minipage}%
  \begin{minipage}[t]{0.495\textwidth}
  \centering
   \includegraphics[width=60mm]{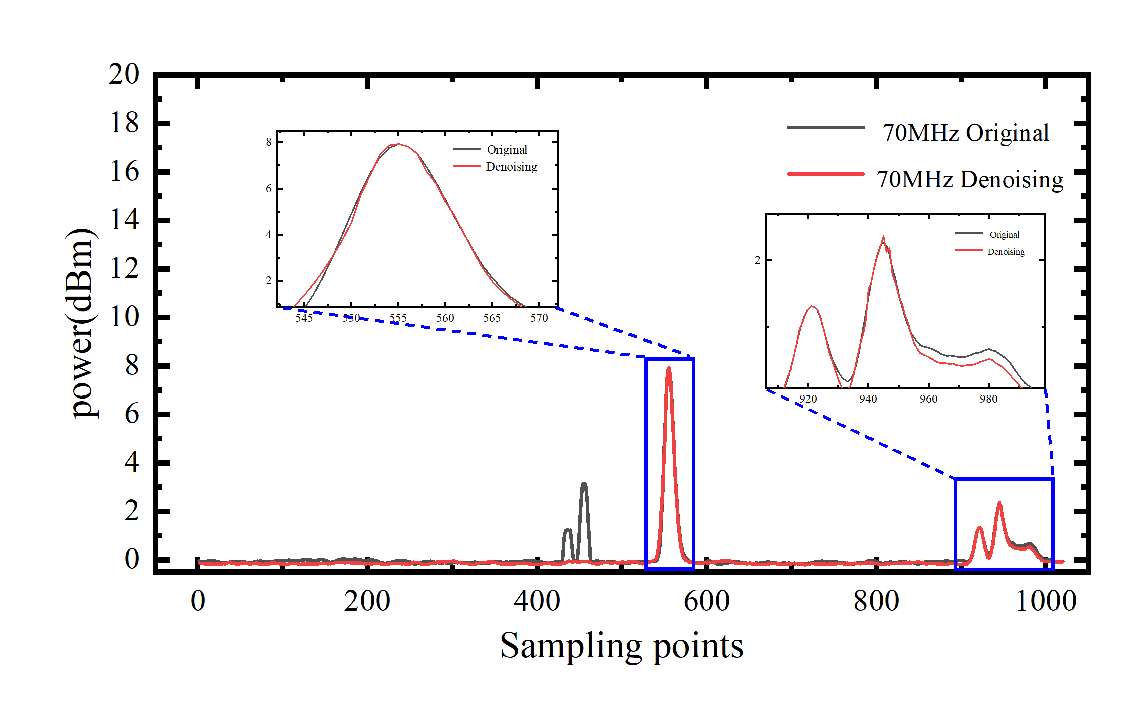}
	  \caption{\label{Fig45}{\small 70MHz spectral line flow.}}
  \end{minipage}
\end{figure}

Currently, the radio analyzer we developed conducts radio observations at three stations: the Qitai Station of the Xinjiang Observatory, the Yunnan Observatory, and the Mianyang Liangdan City Observation Station. The data collected from these three stations are used in this paper to validate the universality and adaptability of the algorithm for 0-100 MHz low-frequency observations.
This paves the way for high-quality data acquisition at more small-scale, low-cost radio observatories in the future.\newline
Through the tests conducted, the noise reduction method based on wavelet transform and mathematical morphology proposed in this paper effectively suppresses noise in individual radio burst events. To further validate the universality of the proposed method, burst cluster events observed at the Qitai Station of the Xinjiang Observatory, the Yunnan Observatory, and the Mianyang Liangdan City Observation Station were selected for verification.
(Fig.~\ref{Fig36}) and (Fig.~\ref{Fig37}) demonstrate that for successive burst events, the algorithm effectively suppresses most narrowband interference, thereby making the burst signals more prominent. Additionally, (Fig.~\ref{Fig38}), (Fig.~\ref{Fig39}), (Fig.~\ref{Fig40}), and (Fig.~\ref{Fig41}) illustrate that the method achieves excellent noise reduction results when applied to single radio burst events at Yunnan Observatory and cluster burst events at Mianyang Liangdan City Observation Station.This confirms the robustness of the method in various observational scenarios.\newline
The above analysis indicates that, due to the similar characteristics of the interfered signals in the time-frequency graph, the algorithm effectively suppresses interference from various locations, such as the Qitai Station of Xinjiang Observatory, Yunnan Observatory, and Liangdan City Observatory in Mianyang. Therefore, it can be preliminarily concluded that the algorithm is highly adaptable and can effectively meet the needs of different observation stations.\newline
\newline
To evaluate the effectiveness of the noise reduction algorithm and its impact on the spectrum signal, the spectral line flow method is used to quantify the intensity and frequency distribution of the signal. The flow of the spectral line reflects the intensity and energy distribution of the signal at a specific frequency. For analysis, the spectral line flow at the same frequency, before and after noise reduction, was extracted from data collected at the Qitai Station of the Xinjiang Observatory on April 18, 2024.
From the spectral line flow analysis in (Fig.~\ref{Fig42}), it is evident that the quality of the spectrum signal data improves significantly after noise reduction in the absence of solar radio bursts.
The interference signal is effectively eliminated, making the spectrum signal smoother and more stable. The peak noise decreases from 0.4 dBm to 0 dBm.
In the spectral line flow displays of (Fig.~\ref{Fig43}), (Fig.~\ref{Fig44}), and (Fig.~\ref{Fig45}),  the intensity of the radio burst spectrum signal is not significantly weakened after noise reduction, but the NBI is effectively suppressed. Furthermore, the flatness of the spectrum of the signal after processing has improved significantly, indicating that the signal is purer and closer to ideal broadband characteristics. This enhancement provides a clearer and more accurate dataset for subsequent analysis and research.\newline
   \begin{figure}
   \centering
   \includegraphics[width=\textwidth, angle=0]{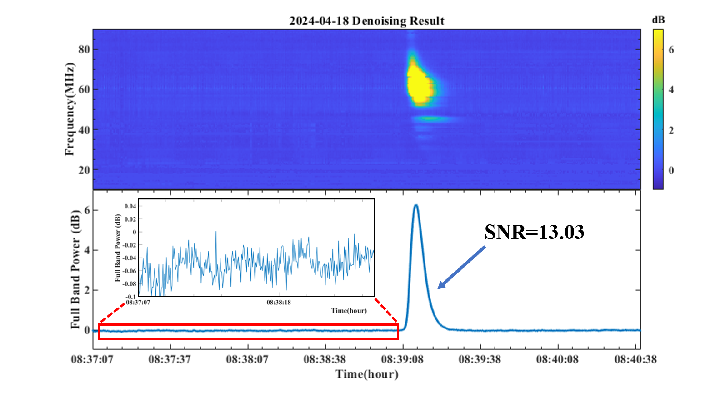}
   \caption{  At the top is the time-frequency spectrogram of the noise-reduced image, while at the bottom is the power change curve over time. }
   \label{Fig46}
   \end{figure}
    \begin{figure}
   \centering
   \includegraphics[width=\textwidth, angle=0]{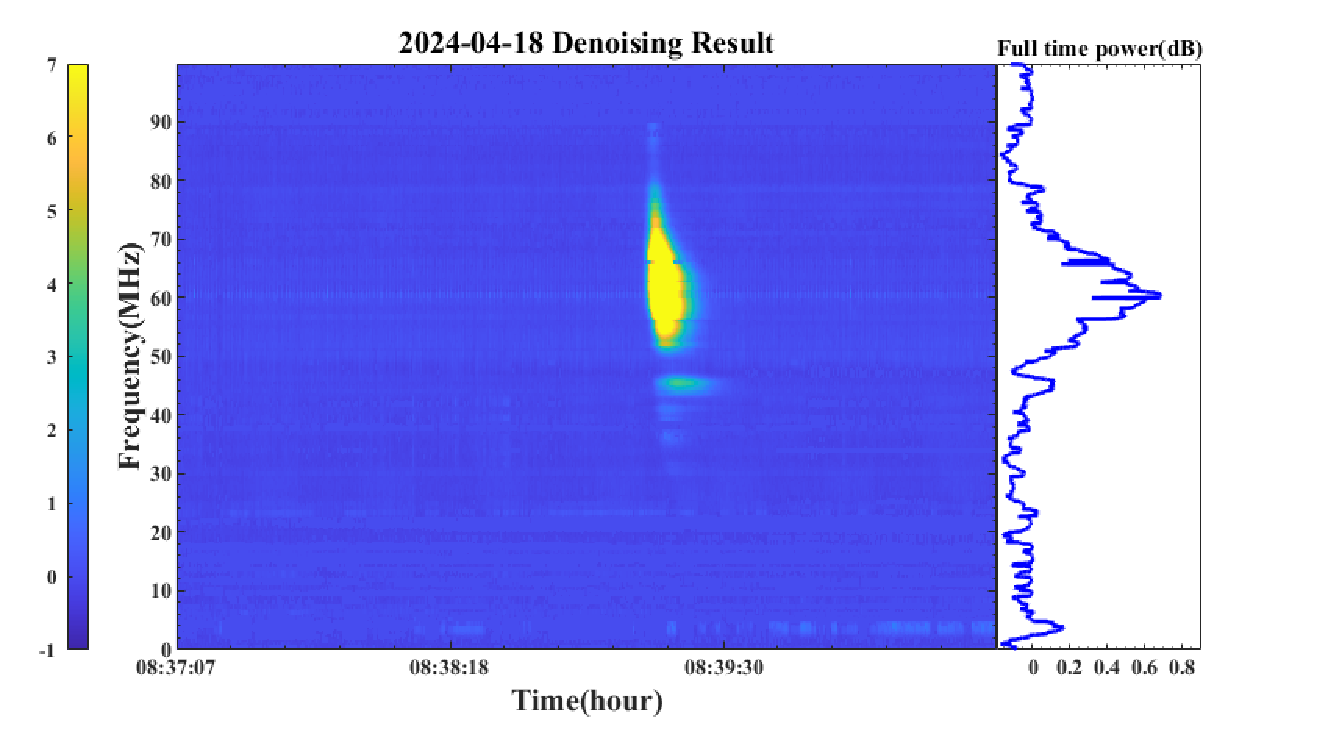}
   \caption{ On the left is the time-frequency spectrogram of the noise-reduced image, and on the right is the full-time power change curve. }
   \label{Fig47}
   \end{figure}
    \begin{figure}
   \centering
   \includegraphics[width=\textwidth, angle=0]{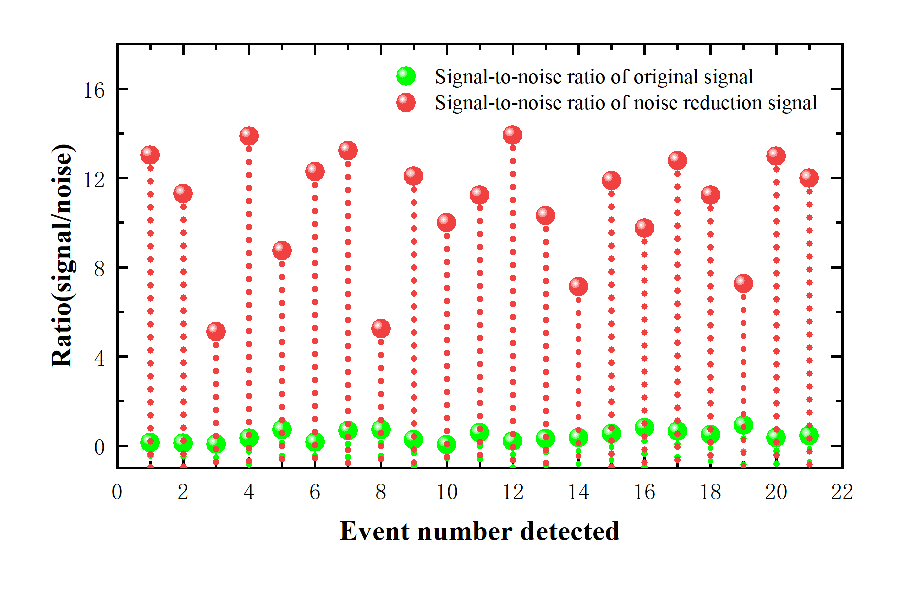}
   \caption{ Signal-to-noise ratio before and after noise reduction for the data recorded in 2024. }
   \label{Fig48}
   \end{figure}
After noise reduction, the signal interference in (Fig.~\ref{Fig46}) and (Fig.~\ref{Fig47}) is effectively suppressed.
The time variation trend of the spectrum signals is restored, and the sharp increase and duration of the solar radio burst spectrum signals can be clearly observed. The signal-to-noise ratio increases from 0.14 to 13.03 dB. The spectral characteristics of the solar radio burst signal are now clearly presented, highlighting the broadband nature of the solar radio burst spectrum.\newline
Additionally, several events observed in 2024 are analyzed in this paper, with the statistical results shown in (Fig.~\ref{Fig48}). The method significantly improves the signal-to-noise ratio, which, in the raw spectrum data, is mostly concentrated between 0 and 1 dB, indicating that the signal is nearly completely obscured by noise. After noise reduction, the signal-to-noise ratio improves significantly, with the mean value reaching 10 dB. This result fully demonstrates the feasibility and effectiveness of the proposed method.
\section{Summary and Prospect}
The presence of RFI significantly hinders the detection and processing of radio spectrum signals. However, most existing methods focus primarily on mitigating RFI, often at the cost of excessive signal loss and a lack of general applicability. To address these issues, this paper proposes a comprehensive RFI suppression method that not only effectively reduces the impact of RFI but also preserves the integrity of the spectrum signals. The method consists of three main steps: data preprocessing, two-dimensional discrete wavelet transform, and mathematical morphology processing.\newline
In this method, most of the background interference in the radio spectrum is removed by data preprocessing. Next, a three-layer wavelet decomposition is applied to the pre-processed signal data using a two-dimensional discrete wavelet transform. The high-frequency and low-frequency wavelet coefficients are then processed using thresholding and absolute median thresholding, respectively. The useful spectrum signal data, post-thresholding, is then separated and reconstructed. Finally, any remaining RFI is further eliminated through mathematical morphology operations, ensuring that useful spectrum signals are preserved while completing the interference suppression process. Compared to traditional RFI suppression methods, this approach places a greater emphasis on protecting useful signals while effectively suppressing RFI. Experimental results based on measured solar radio burst spectrum signal data demonstrate that this method successfully preserves useful signals while mitigating RFI. However, it is important to highlight that safeguarding useful signals during RFI suppression is just as crucial as the RFI reduction itself. Failure to do so may lead to the loss of valuable spectrum information, thus reducing the resolution of the radio spectrum image. The method proposed in this paper combines the advantages of wavelet transforms for multi-scale analysis with the ability of mathematical morphology operations to extract morphological features. The authors believe that this method can be a valuable tool for suppressing RFI in single-station radio astronomy observations.
\section{Acknowledgments}
The authors sincerely thank all the reviewers and editors for their valuable comments and suggestions on this paper. Your feedback has greatly contributed to making the content more rigorous and refined.\newline
 This research was funded by the National Key Research and Development Program's intergovernmental International Science and Technology Innovation Cooperation project, titled Remote Sensing and Radio Astronomy Observation of Space Weather in Low and Middle Latitudes (project number: 2022YFE0140000).\newline
Supported by International Partnership Program of Chinese Academy of Sciences. Grant No. 114A11KYSB20200001

\newpage
\bibliographystyle{raa}
\small\bibliography{bibtex}

\label{lastpage}

\end{document}